\documentclass[useAMS,usenatbib]{mn2e}
\usepackage{amssymb}
\usepackage{graphicx}
\usepackage{amsmath}
\usepackage{natbib}
\usepackage{epsfig}
\usepackage{epstopdf}
\newcommand{\asec} {\mbox{$^{\prime \prime}$}}

\usepackage{mathrsfs}
\usepackage{amsfonts}

\title[Radio-Loudness in XMM-COSMOS]{Inter-comparison of
Radio-Loudness Criteria for Type 1 AGNs in the XMM-COSMOS Survey}

\author[Heng Hao et al.]{Heng Hao$^{1,2}$\thanks{E-mail:henghao@post.harvard.edu}, Mark T. Sargent$^{3,4,5}$,
Martin Elvis$^{2}$, Eva Schinnerer$^{5}$, Gianni
\newauthor
Zamorani$^{6}$, Luis C. Ho$^{7,8,9}$, Jennifer L. Donley$^{10}$,
Francesca Civano$^{2,11,12}$, Vernesa
\newauthor
Smol\v{c}i\'{c}$^{13}$, Annalisa Celotti$^{1,14,15}$, Joanna
Kuraszkiewicz$^{2}$, Mara Salvato$^{16,17}$,
\newauthor
 Marcella Brusa$^{6,18,19}$, Peter Capak$^{20}$,
Chris L. Carilli$^{21}$, Andrea Comastri$^{6}$,
\newauthor
Chris D. Impey$^{22}$, Knud Jahnke$^{5}$, Anton M. Koekemoer$^{23}$,
Kevin Schawinski$^{24}$,
\newauthor
Jonathan R. Trump$^{25}$, C. Megan Urry$^{11}$, Cristian Vignali$^{6,18}$, Min Yun$^{26}$\\
$^{1}$SISSA, Via Bonomea 265, I-34136 Trieste, Italy\\
$^{2}$Harvard-Smithsonian Center for Astrophysics,
60 Garden Street, Cambridge, MA 02138, USA\\
$^{3}$Astronomy Center, Department of Physics and Astronomy,
University of Sussex, Brighton BN1 9QH, UK\\
$^{4}$CEA-Saclay, Service d'Astrophysique, Orme des Merisiers, Bat.
709, 91191 Gif-sur-Yvette, France\\
$^{5}$Max-Planck-Institut f\"ur Astronomie, K\"onigstuhl 17,
Heidelberg, D-69117, Germany\\
$^{6}$INAF - Osservatorio Astronomico di Bologna, via Ranzani 1,
I-40127 Bologna, Italy\\
$^{7}$Kavli Institute for Astronomy and Astrophysics, Peking University, Beijing 100871, China\\
$^{8}$Department of Astronomy, Peking University, Beijing, China \\
$^{9}$The Observatories of the Carnegie Institute for
Science, Santa Barbara Street, Pasadena, CA 91101, USA\\
$^{10}$Los Alamos National Laboratory, Los Alamos, NM 87544, USA\\
$^{11}$Yale Center for Astronomy and Astrophysics, 260 Whitney ave, New Haven, CT 06520, USA\\
$^{12}$Dartmouth College, Department of Physics and Astronomy, 6127
Wilder Lab, Hanover, NH 03755\\
$^{13}$Physics Department, University of Zagreb, Bijeni\v{c}ka cesta
32, 10002 Zagreb, Croatia\\
$^{14}$INAF - Osservatorio Astronomico di Brera, via E. Bianchi 46,
I-23807 Merate, Italy\\
$^{15}$INFN - Sezione di Trieste, via Valerio 2, 34127, Trieste,
Italy\\
$^{16}$IPP - Max-Planck-Institute for Plasma Physics, Boltzmann
Strasse 2, D-85748, Garching bei M\"{u}nchen, Germany\\
$^{17}$Excellence Cluster, Boltzmann Strasse 2, D-85748,
Garching bei M\"{u}nchen, Germany\\
$^{18}$Dipartimento di Fisica e Astronomia, Universit\`{a} degli
studi di Bologna, viale Berti Pichat 6/2 40127 Bologna, Italy\\
$^{19}$Max Planck Institute f\"ur Extraterrestrische Physik,
Postfach 1312, 85741, Garching bei M\"{u}nchen, Germany\\
$^{20}$Spitzer Science Center, MS 314-6, California Institute of
Technology, Pasadena, CA 91125, USA\\
$^{21}$ National Radio Astronomy Observatory, P.O. Box 0, Socorro,
NM 87801, USA\\
$^{22}$Steward Observatory, University of Arizona, 933 North Cherry
Avenue, Tucson, AZ 85721, USA\\
$^{23}$Space Telescope Science Institute, 3700 San Martin Drive,
Baltimore, MD 21218, USA\\
$^{24}$Institute for Astronomy, Department of Physics, ETH
Zurich, Wolfgang-Pauli-Strasse 16, CH-8093 Zurich, Switzerland\\
$^{25}$Department of Astronomy and Astrophysics, Pennsylvania State
University, University Park, PA 16802, USA\\
$^{26}$Department of Astronomy, University of Massachusetts,
Amherst, MA 01003, USA}

\begin{document}

\date{Version Jun 1st, 2014.}

\pagerange{\pageref{firstpage}--\pageref{lastpage}}\pubyear{2014}

\maketitle

\label{firstpage}

\begin{abstract}
Limited studies have been performed on the radio-loud fraction in
X-ray selected type 1 AGN samples. The consistency between various
radio-loudness definitions also needs to be checked. We measure the
radio-loudness of the 407 type 1 AGNs in the XMM-COSMOS quasar
sample using nine criteria from the literature (six defined in the
rest-frame and three defined in the observed frame):
$R_L=\log(L_{5GHz}/L_B)$, $q_{24}=\log(L_{24\mu m}/L_{1.4GHz})$,
$R_{uv}=\log(L_{5GHz}/L_{2500\AA})$, $R_{i}=\log(L_{1.4GHz}/L_i)$,
$R_X=\log(\nu L_{\nu}(5GHz)/L_X)$,
$P_{5GHz}=\log(P_{5GHz}(W/Hz/Sr))$, $R_{L,obs}=\log(f_{1.4GHz}/f_B)$
(observed frame), $R_{i,obs}=\log(f_{1.4GHz}/f_i)$ (observed frame),
and $q_{24, obs}=\log(f_{24\mu m}/f_{1.4GHz})$ (observed frame).
Using any single criterion defined in the rest-frame, we find a low
radio-loud fraction of $\lesssim$5\% in the XMM-COSMOS type 1 AGN
sample, except for $R_{uv}$. Requiring that any two criteria agree
reduces the radio-loud fraction to $\lesssim 2\%$ for about 3/4 of
the cases. The low radio-loud fraction cannot be simply explained by
the contribution of the host galaxy luminosity and reddening. The
$P_{5GHz}=\log(P_{5GHz}(W/Hz/Sr))$ gives the smallest radio-loud
fraction. Two of the three radio-loud fractions from the criteria
defined in the observed frame without k-correction ($R_{L,obs}$ and
$R_{i,obs}$) are much larger than the radio-loud fractions from
other criteria.
\end{abstract}

\begin{keywords}
galaxies: evolution; quasars: general; surveys
\end{keywords}

\section{Introduction}

Quasars are often classified into radio-loud (RL) and radio-quiet
(RQ), based on the presence or absence of strong radio emission.
Radio-loud quasars are generally some three orders of magnitude more
powerful at GHz radio frequencies relative to their optical or
infrared fluxes than RQ quasars, while in the rest of their spectral
energy distributions (SEDs), from mid-infrared to X-ray, there are
only subtle differences between them (e.g., Elvis et al., 1994, E94
hereinafter). The strong radio emission is a result of RL quasars
having a relativistic jet that generates synchrotron radiation in
the radio (see review by Harris \& Krawczynski 2006).

However, even RQ quasars can be detected as radio sources
(Kellermann et al. 1989). This has led to two opposing views of the
radio-loudness distribution which have long been debated. The first
is that the radio-loudness distribution is bimodal (e.g. Kellermann
et al. 1989; Miller et al. 1990; Visnovsky et al. 1992; Ivezi\'{c}
et al. 2002). The other is that the distribution is continuous with
no clear dividing line (Cirasuolo et al. 2003).

Typically, $\sim$10\% of all quasars in optically selected samples
are RL (e.g. Kellermann et al. 1989; Urry \& Padovani 1995;
Ivezi\'{c} et al. 2002). Here we examine the RL fraction of the 413
type 1 AGNs in the XMM-COSMOS sample to check both the RL fraction
in X-ray selected samples and the consistency among various
radio-loudness criteria. In this paper, we adopt the WMAP 5-year
cosmology (Komatsu et al., 2009), with H$_0$
=71~km~s$^{-1}$~Mpc$^{-1}$, $\Omega_M$ = 0.26 and
$\Omega_{\Lambda}$= 0.74.

\section{COSMOS Type 1 AGN Photometry} \label{s:sample}

The XMM-COSMOS survey (Hasinger et al. 2007) detected 1848 point
sources down to $\sim10^{-15}$~erg~cm$^{-2}$s$^{-1}$. Using a
likelihood ratio technique, Brusa et al (2007, 2010) identified
unique counterparts of 1577 (85\%) XMM-COSMOS sources in the optical
photometric catalog (Capak et al. 2007). A total of 886 XMM-COSMOS
sources ($\sim$50\%) have well-determined spectroscopic redshifts
from optical spectra (Trump et al. 2009a, Schneider et al. 2007,
Lilly et al. 2007, 2009). From these spectra, 413 are identified as
type 1 AGN, with emission line FWHM$>2000~$km s$^{-1}$, forming the
XMM-COSMOS type 1 AGN sample (XC413, Elvis et al. 2012, hereafter
Paper I). The XC413 sample has 43 photometry bands extending from
radio to X-ray, and spans a large redshift range ($0.1\leq z
\leq4.3$, with median 1.6), as well as both large apparent magnitude
($16.8\leq i_{AB}\leq25.0$ with median 21.2) and intrinsic
luminosity ($44.3\leq\log L_{bol}\leq47.4$ with median 45.7) ranges.
We now briefly review the optical, infrared (IR) and radio flux
measurements which are crucial to the analysis.

All 413 quasars in XC413 have B band and J band detections (Paper
I). In the mid-infrared range, 385 detections were obtained in the
S-COSMOS MIPS $24\mu m$ imaging (Sanders et al. 2007; Le Floc'h et
al. 2009) by searching for counterparts within 2\asec\ of the
optical counterpart. With the exception of one source (located
outside the $24\mu m$ survey area), we derived 3$\sigma$ upper flux
density limits ($\sim$54$\mu$Jy on average) for the remaining 27
unmatched quasars using a coverage-based rms map.

In the radio, the VLA-COSMOS 1.4 GHz survey detected 2865 sources at
S/N=5 (rms=8-12 $\mu$Jy, depending on the position in the field;
Schinnerer et al. 2010). Radio counterparts to the XMM X-ray sources
were determined by cross-correlating the optical quasar positions to
source positions in the VLA-COSMOS joint catalog (Schinnerer et al.
2010) within a radius of 1\asec. This resulted in 61 (15\% of 413)
successful matches with the XC413 sample.

For the unmatched XMM-COSMOS quasars, the AIPS/MAXFIT peak finding
algorithm was used to search for additional radio detections within
a 2.5\asec$\times$2.5\asec\ box centered on the optical coordinates.
The box size is chosen because the resolution of the radio beam is
2.5\asec. This yielded 78 additional detections in the
3$\sigma$-5$\sigma$ range. In all, we have 139 sources with larger
than 3$\sigma$ radio detection. We computed their total flux
assuming that they are unresolved at 1.4GHz (beam FWHM 2.5\asec).
For lower significance peaks (286 out of 413) we adopted 3$\sigma$
upper flux density limits based on the local rms noise (calculated
within a 17.5\asec$\times$17.5\asec\ box) at the position of the
radio source. This box size was chosen based on the tests we made to
obtain the most accurate map for the VLA-COSMOS Deep Project mosaic.

In summary, out of the 413 XMM-COSMOS quasars, 407 have either
$>3\sigma$ VLA detections or upper limits. We have no radio flux
information about the remaining 6 AGNs as they lie outside the
VLA-COSMOS 1.4 GHz coverage area, and in one case also outside the
MIPS-COSMOS 24$\mu$m coverage area. We will only discuss the radio
loudness for these 407 type 1 AGNs (XC407 sample) in this paper.

\section{Radio-Loudness Definitions} \label{s:radio}

\begin{figure*}
\includegraphics[angle=0,width=0.32\textwidth]{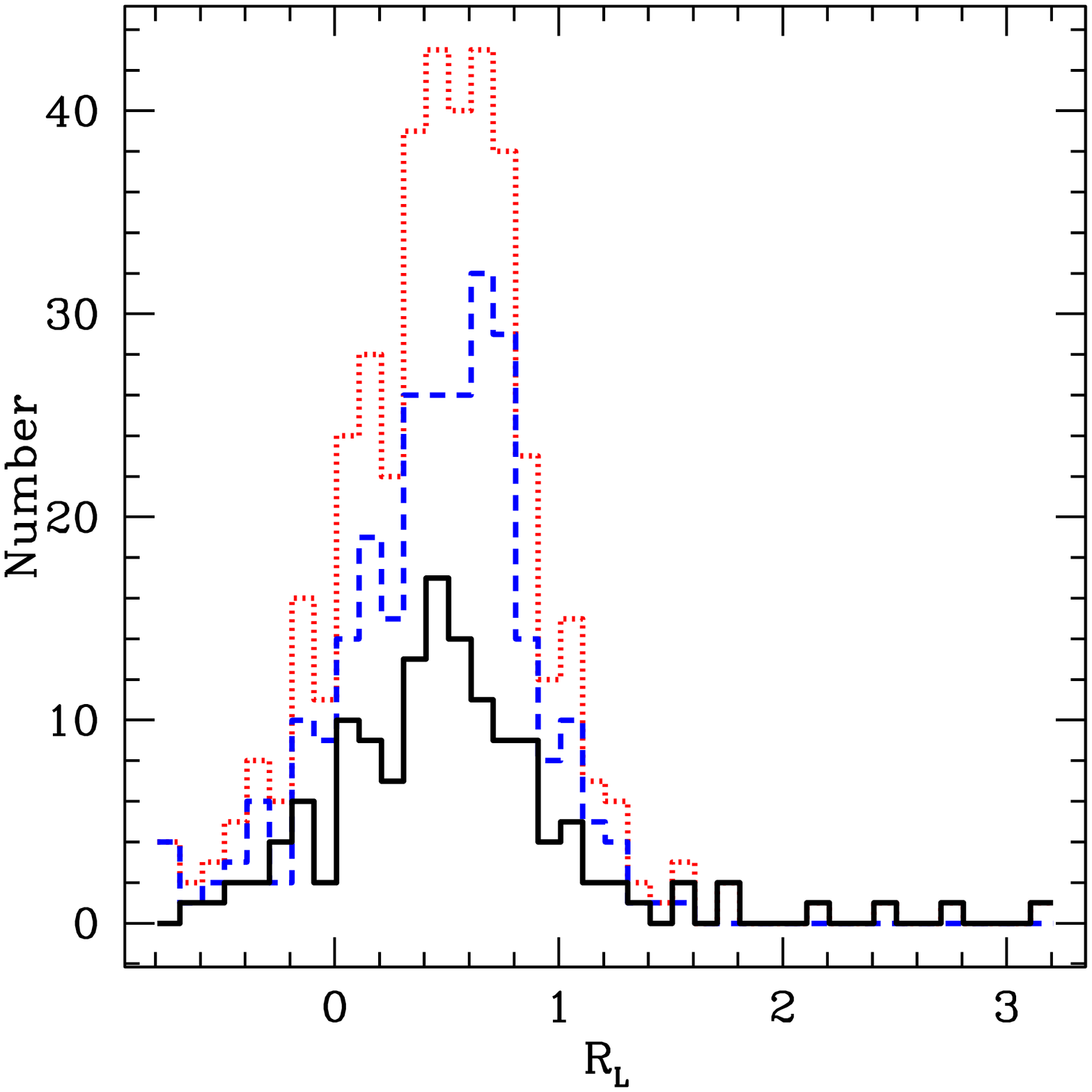}
\includegraphics[angle=0,width=0.32\textwidth]{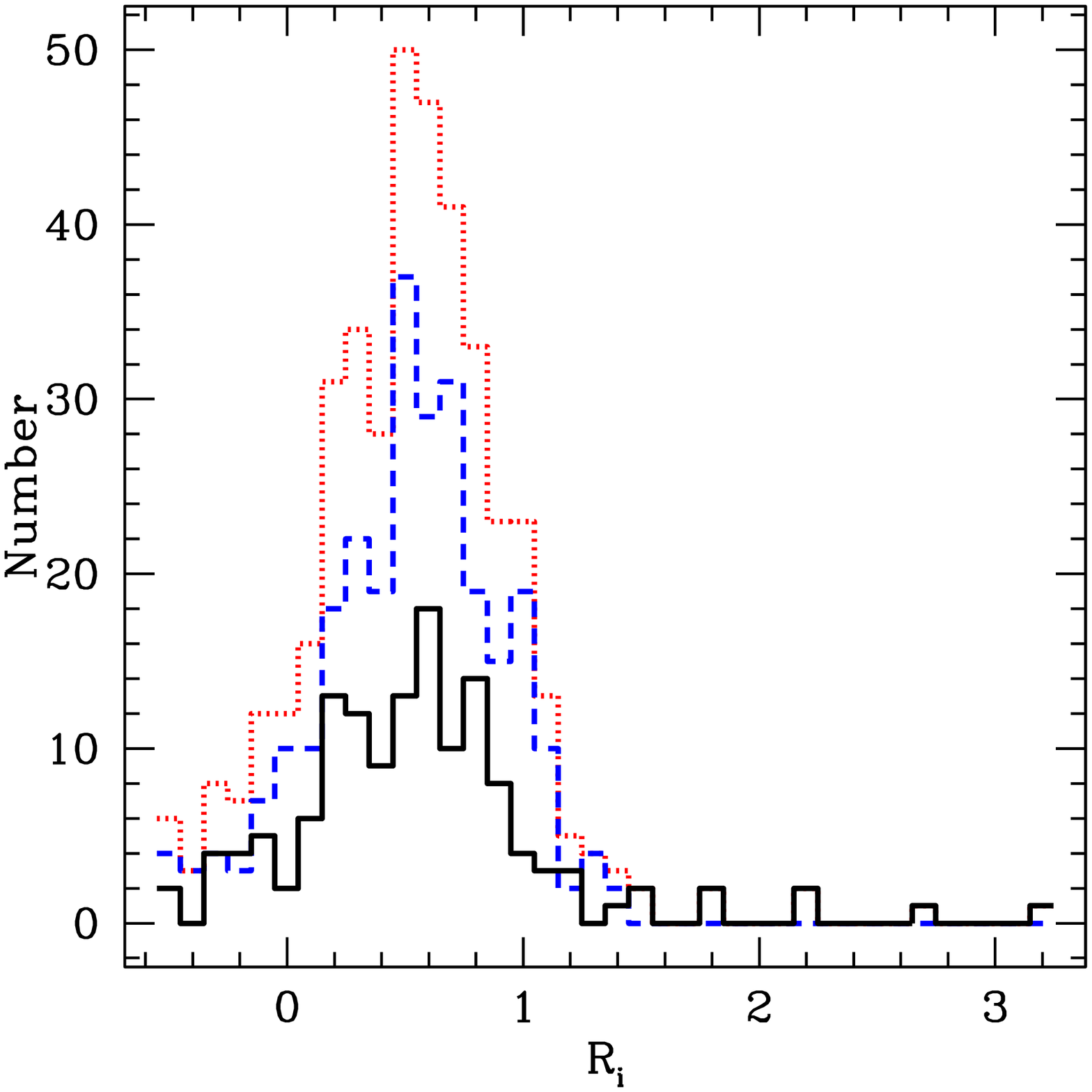}
\includegraphics[angle=0,width=0.32\textwidth]{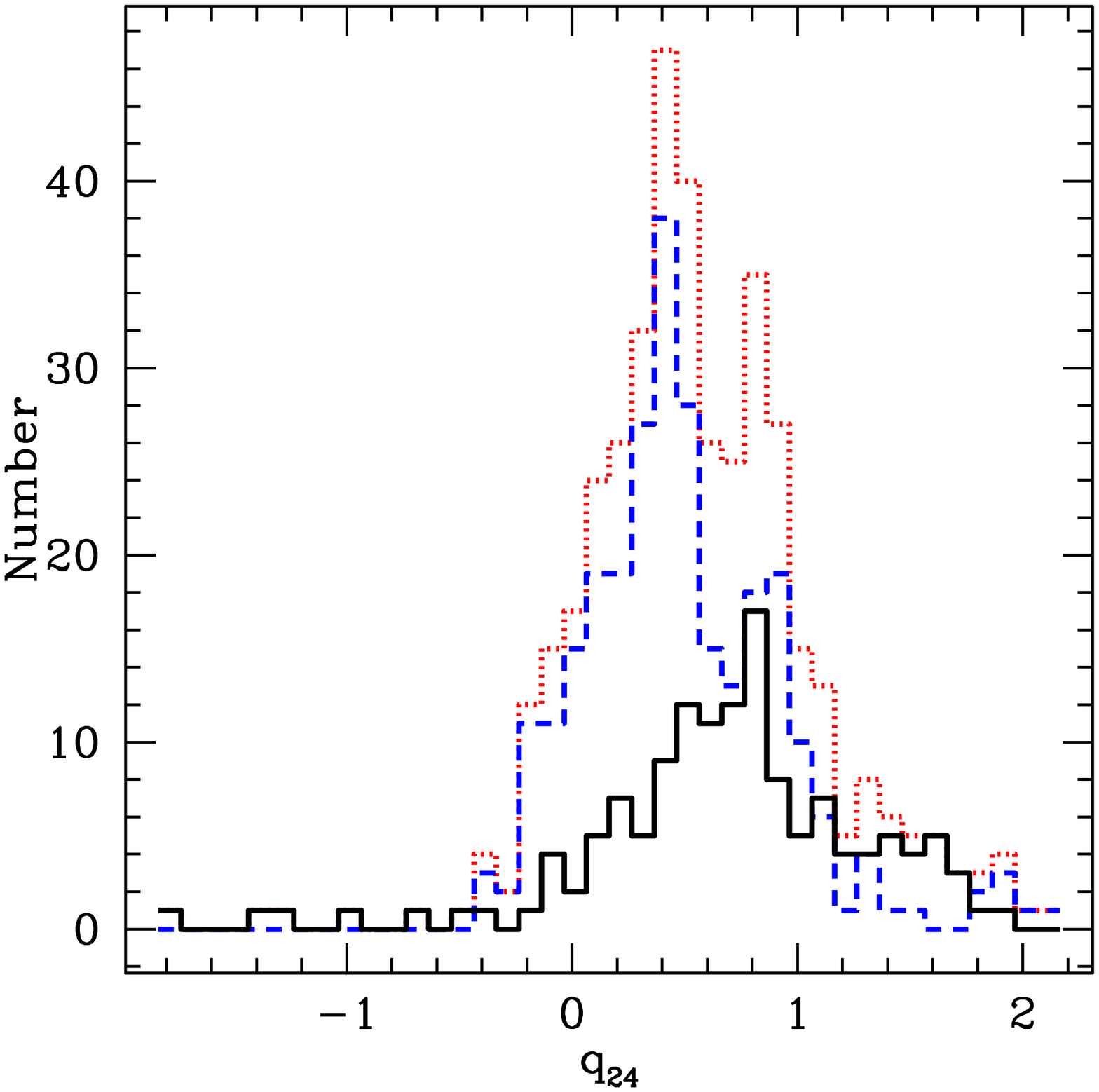}
\includegraphics[angle=0,width=0.32\textwidth]{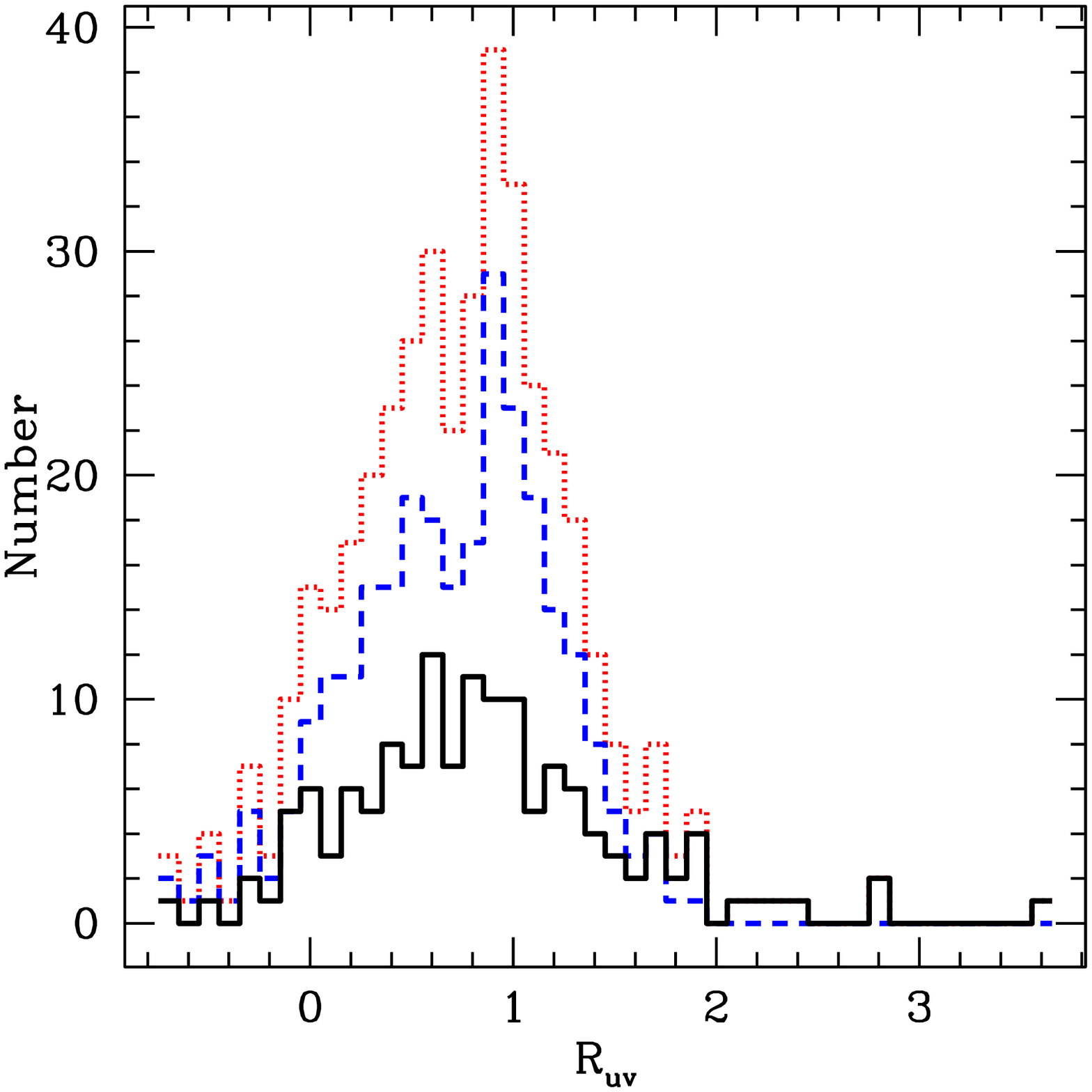}
\includegraphics[angle=0,width=0.32\textwidth]{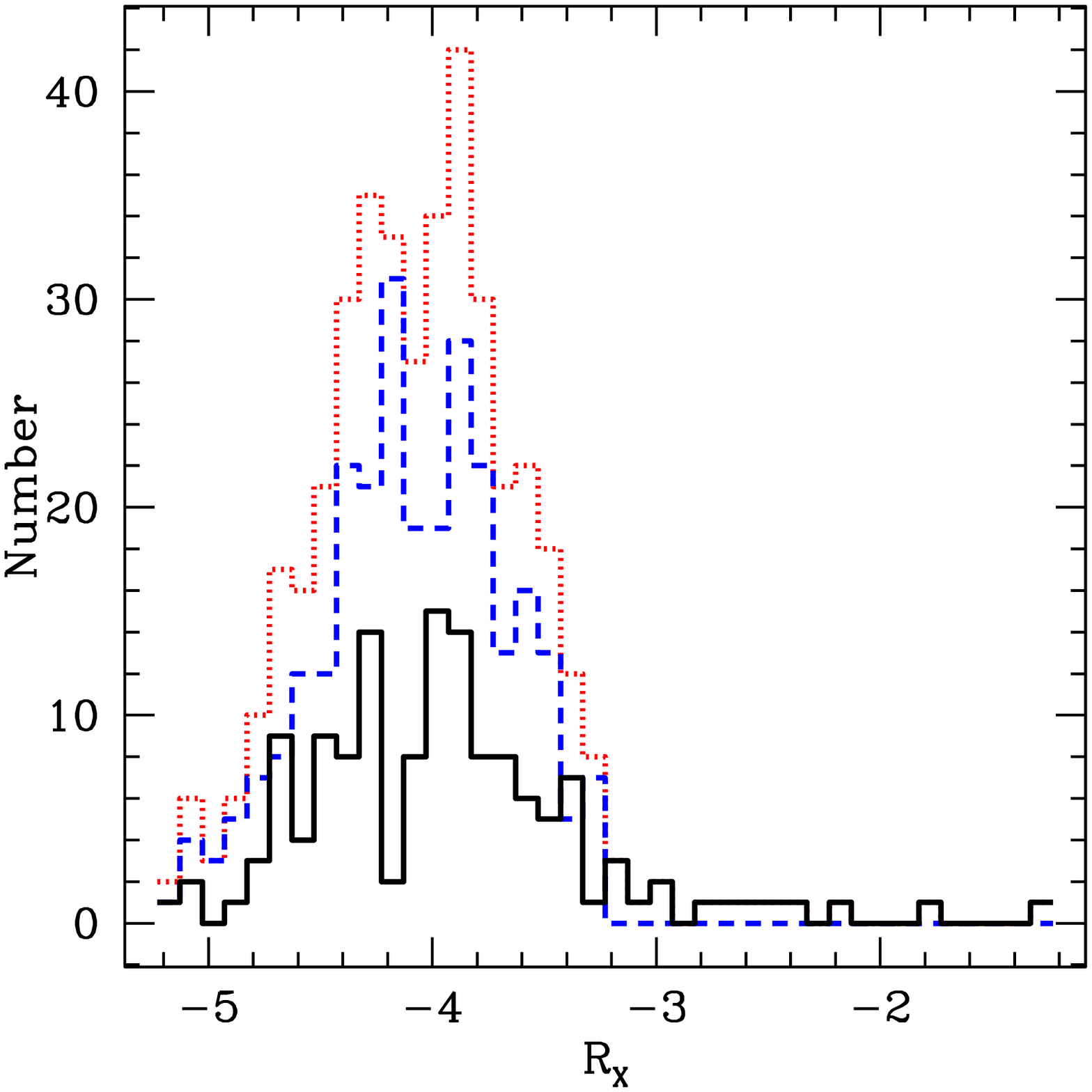}
\includegraphics[angle=0,width=0.32\textwidth]{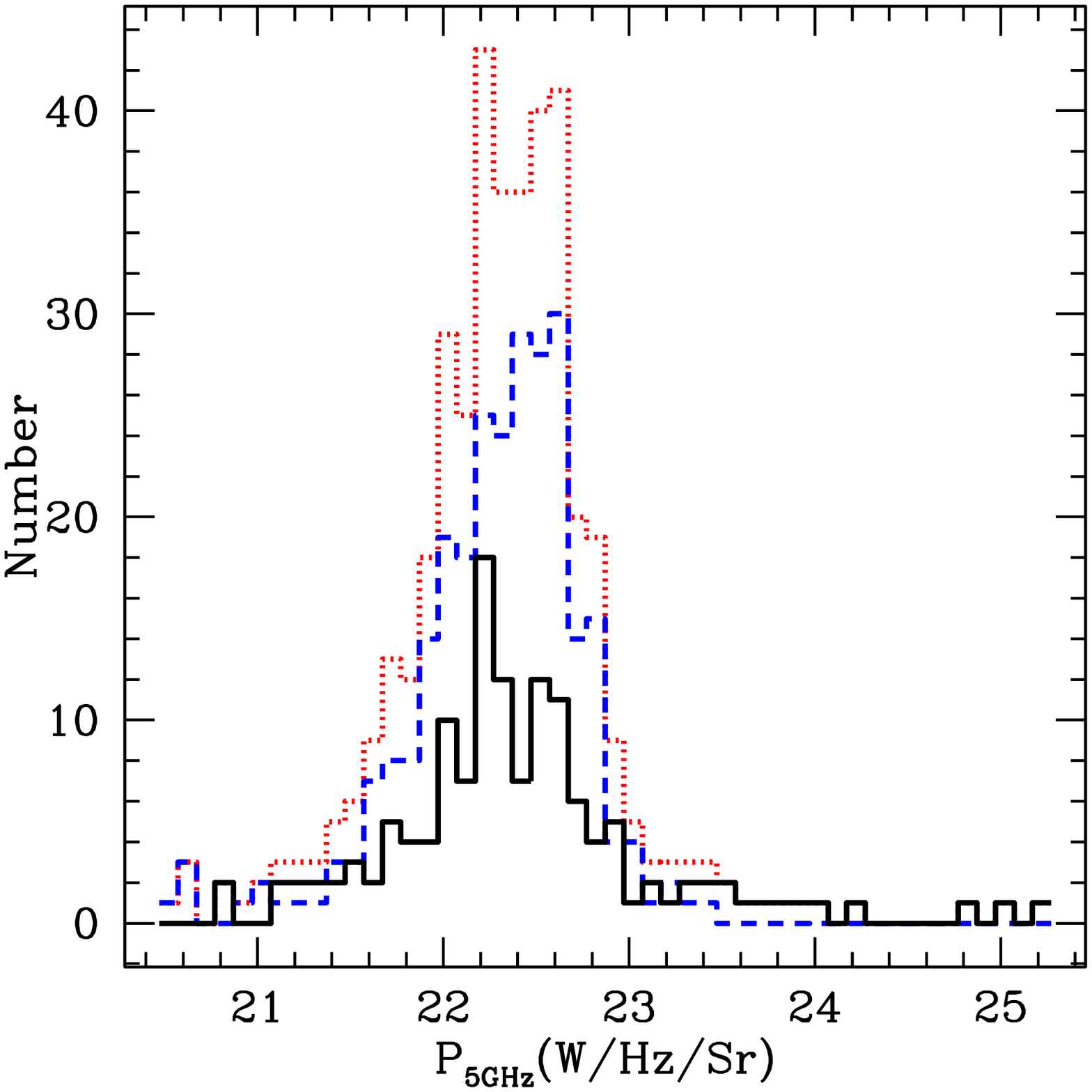}
\includegraphics[angle=0,width=0.32\textwidth]{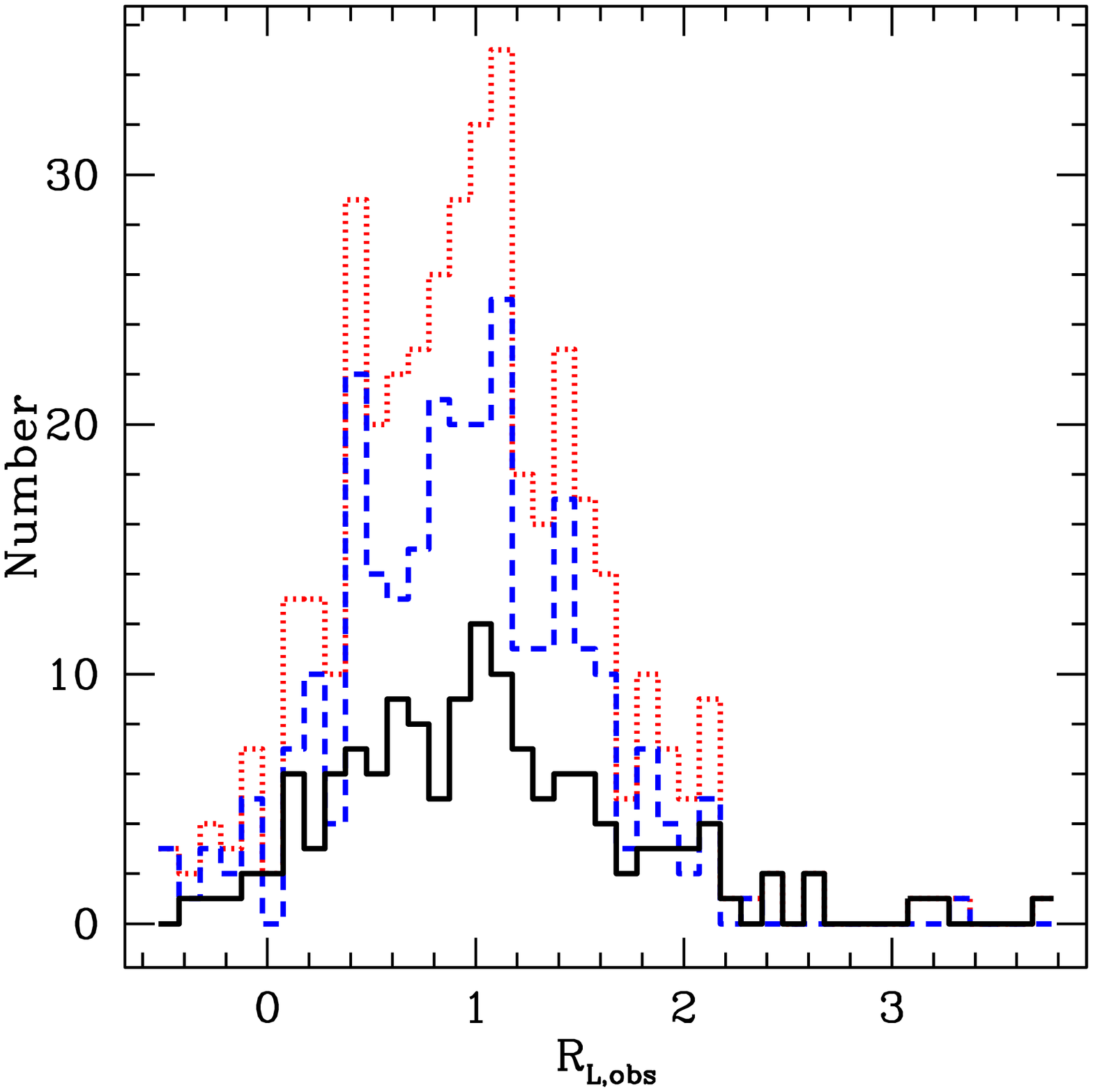}
\includegraphics[angle=0,width=0.32\textwidth]{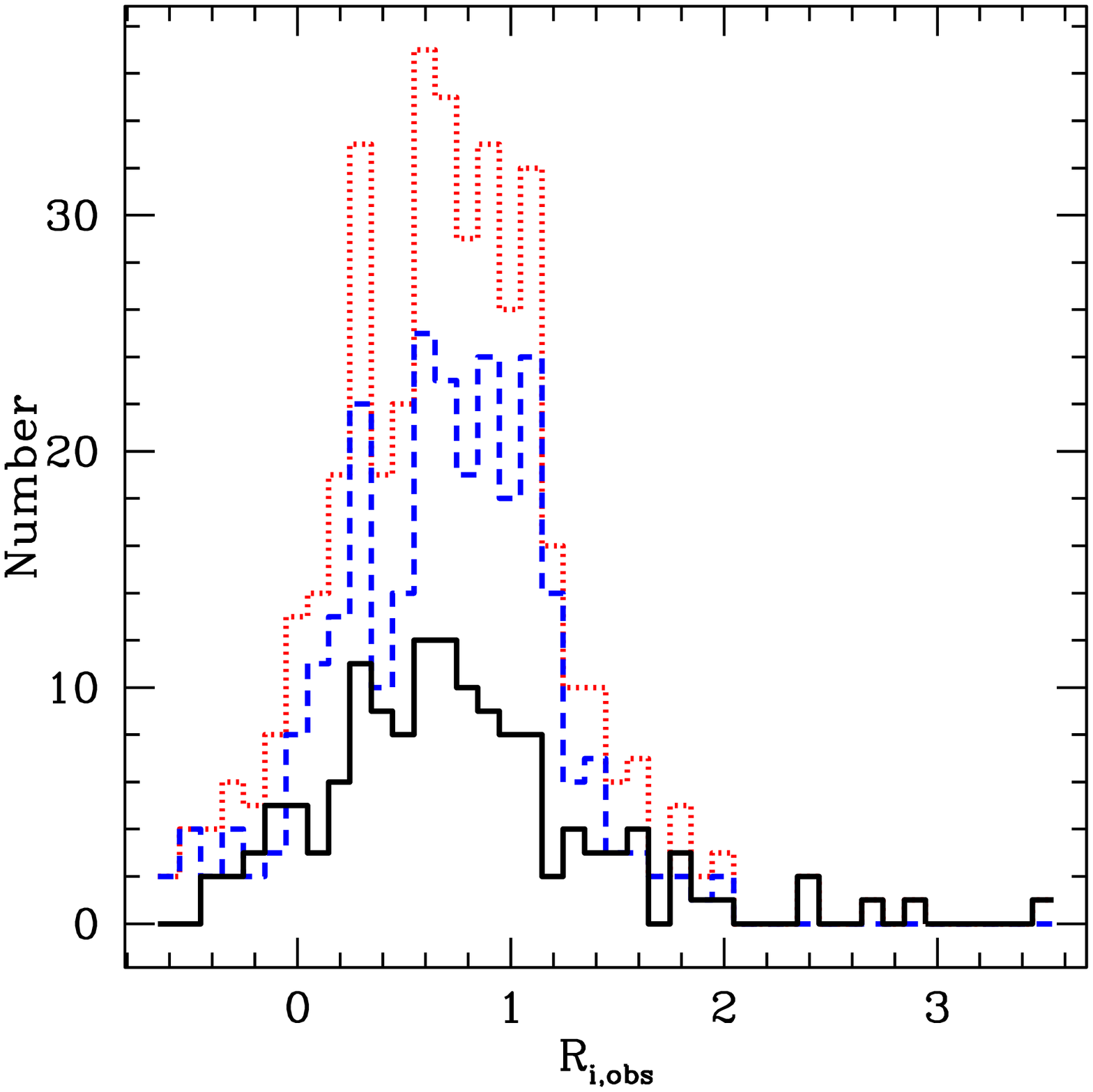}
\includegraphics[angle=0,width=0.32\textwidth]{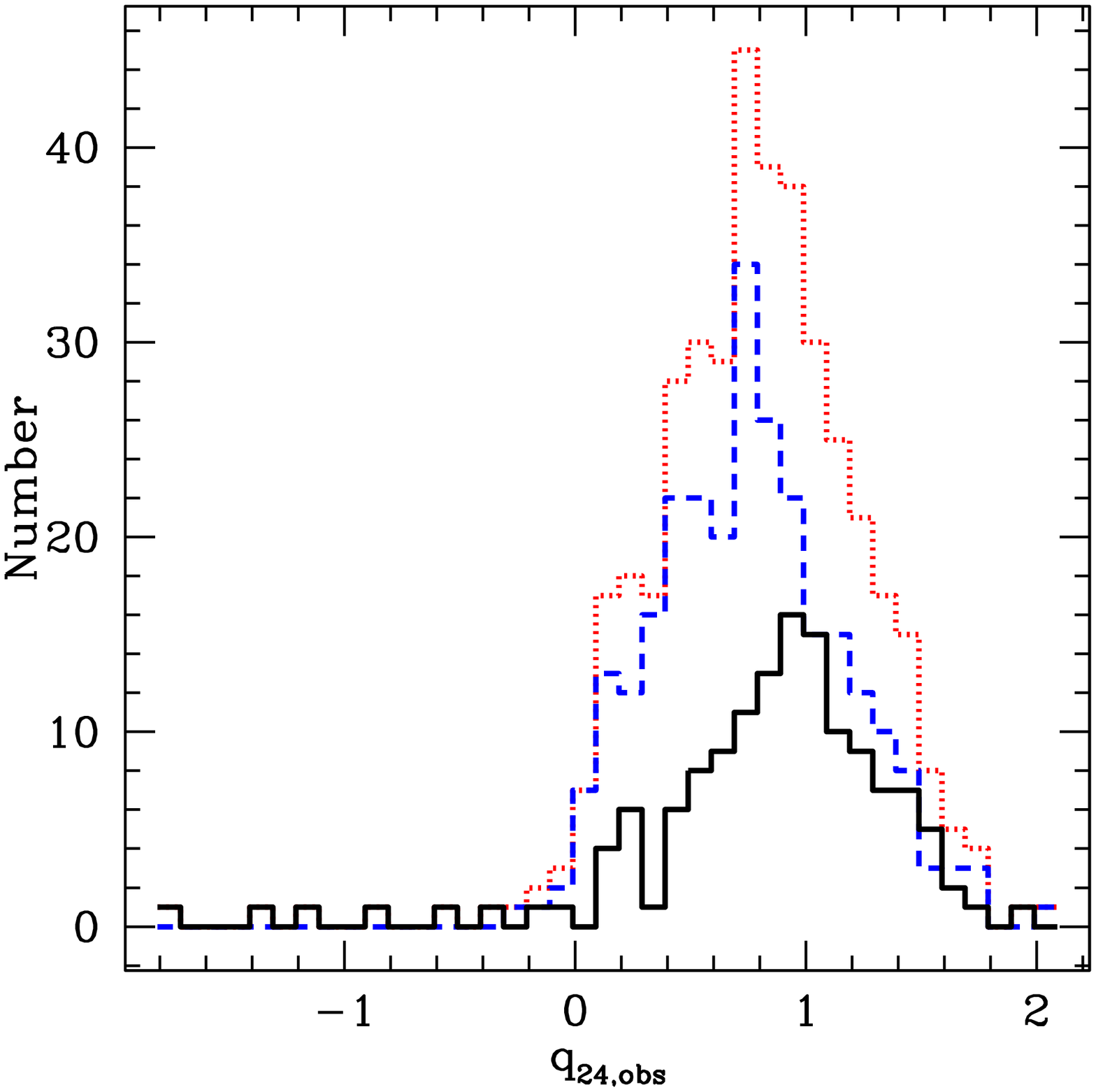}
\caption{Distribution of radio-loudness measures: $R_L$, $R_i$,
$q_{24}$, $R_{uv}$, $R_X$, $P_{5GHz}$, $R_{L,obs}$, $R_{i,obs}$, and
$q_{24,obs}$. The black solid line show the distribution for quasars
with radio detections; the blue dashed line show the distribution
for quasars with upper limits in radio; the red dotted line show the
distribution for all the 407 quasars with radio detection or upper
limits. \label{rlhist}}
\end{figure*}

Several criteria have been used to classify quasars as RL or RQ. As
noted, mere radio detection is not enough. Radio power, either alone
or relative to some other band is typically used. We determined the
radio-loudness of the XC407 for nine definitions currently in use in
the literature:

\begin{enumerate}

\item $R_L$: the luminosity ratio of radio to optical emission
$R_L = \log(L_{5GHz}/L_{B})$ (Wilkes \& Elvis 1987, Kellermann et
al. 1989), with $R_L>$1 defining a RL source. This logarithmic (base
ten, as for all following definitions) radio-to-optical luminosity
ratio is the most widespread criterion for RL. COSMOS does not have
5~GHz coverage, but, as most of the sample sources are at redshift
1--2 (Paper I), the observed 1.4~GHz VLA band is close to the
emitted 5~GHz frequency. We converted the observed 1.4 GHz
luminosity to a rest-frame 5 GHz luminosity by assuming
$f_{\nu}\propto \nu^{-0.5}$ (e.g., Ivezi\'{c} 2004). For most of the
quasars in the XC407 sample at redshift $z\sim2$, the observed 1.4
GHz is at rest-frame $\sim 4.2$ GHz, the residual k-correction is
only 9\%. The B band luminosity is the luminosity at rest-frame B
band ($\lambda_{eff}=4483\AA$) retrieved from the rest-frame SED for
each quasar, that is the linear interpolation of the adjacent
observed photometry after moving to the rest-frame.

\item $R_{L,obs}$: We also calculated
$R_{L,obs}=\log(f_{1.4GHz}/f_B)$ in the observed frame without
k-correction for comparison, with $R_{L,obs}>1$ defined as
radio-loud. This criteria is typically used when redshift
information is not available.

As most of the XMM-COSMOS AGNs lie at redshift 1--2, the 1.4 GHz
band lies close to rest-frame 5 GHz and the J-band lies close to
rest-frame B band. Hence, for a consistency check, we can adopt
$R_{J} =\log(f_{1.4GHz}/f_{J})$ in the observed frame as an
alternative definition to RL that does not involve any assumptions
about the k-correction.

\item $R_i$: Balokovi\'{c} et al. (2012) defined the radio-loudness as
the radio to i band luminosity ratio: $R_i=\log(L_{5GHz}/L_i)$. Here
we calculate $R_i$ using the same k-correction in the radio as in
$R_L$ and retrieve the rest-frame i band ($\lambda_{eff}=7523\AA$)
luminosity from the rest-frame SED by interpolation. We define
$R_i>1$ as radio-loud.

\item $R_{i, obs}$: Ivezi\'{c} et al. (2002) defined the radio-loudness as
$R_{i, obs}=\log(f_{1.4GHz}/f_i)$ in the observed frame without
k-correction and $R_{i,obs}>1$ was defined as radio-loud.
Considering most of the XMM-COSMOS AGNs lie at redshift 1--2, the
1.4 GHz band lies close to rest-frame 5 GHz and the K-band lies
close to rest-frame i band. Similarly, we can adopt $R_{K}
=\log(f_{1.4GHz}/f_{K})$ in the observed frame without k-correction
as an alternative definition to radio-loudness.

\item $q_{24}$: Appleton et al. (2004) introduced a new definition
of radio-loudness using the Spitzer/MIPS 24~$\mu$m flux: $q_{24}$ =
$\log(L_{24\mu m}/L_{1.4GHz})$, with $q_{24}<0$ defined as RL. We
calculated $q_{24}$ in the rest-frame assuming the same power law in
the radio as for the $R_L$ definition and the rest frame
24$\rm{\mu}$m retrieved from the rest-frame SED of each quasar.
Kuraszkiewicz et al. (in preparation) studied the correlation
between $q_{24}$ and $R_L$ for a sample with limited reddening or
host contamination, and find that, on average, $R_L >1$ corresponds
to $q_{24} < 0.24$, rather than $q_{24} < 0$. We use $q_{24,K}$ to
represent the criterion using this alternative dividing line.

\item $q_{24, obs}$: For easy comparison with $R_{L, obs}$, which is
defined in the observed frame, we also check $q_{24}$ in the
observed frame and define it as $q_{24, obs}=\log(f_{24\mu
m}/f_{1.4GHz})$.

\medskip
\medskip

Out of the 407 quasars, 25 have upper flux density limits both at
1.4 GHz and 24 $\mu$m. For these sources, $q_{24}$ and $q_{24, obs}$
cannot be determined and we thus excluded these 25 sources from the
discussion of the results for $q_{24}$ and $q_{24, obs}$.

\medskip
\medskip

\item $R_{uv}$: Stocke et al. (1992) used the ratio between the 2500\AA\
UV luminosity and radio luminosity:
$R_{uv}=\log(L_{5GHz}/L_{2500\AA})$ (see also Jiang et al. 2007).
This criterion has the advantage of using the peak of the SED to
define radio-loudness, but is strongly affected even by modest
reddening. For example, $E(B-V)>0.3$ decreases the 2500\AA\ UV
luminosity by a factor $>7$. The effect is to make more objects
appear RL in X-ray selected samples, such as the XC407 sample, as
these include a large number of sources with significant optical
reddening compared to optically-selected samples (Paper I, Hao et
al. 2013, 2014). This criterion is therefore less useful for X-ray
selected samples. In the \emph{Einstein} Extended Medium Sensitivity
Survey (EMSS), radio-loudness is defined with
$\alpha_{ro}=R_{uv}/5.38$ (Della Ceca et al. 1994, Zamorani et al.
1981). Sources with $\alpha_{ro}>0.35$ are defined as RL, that is
equivalent to $R_{uv}>1.88$. We use $R_{uv,D}$ to represent the
criterion.

\item $R_{X}$: Terashima \& Wilson (2003) proposed a criterion based
on the ratio between luminosity at 5 GHz and X-ray luminosity in the
$2-10~keV$ band. Sources with $R_X=\log(\nu L_{\nu} (5GHz)/L_X)>-3$
are RL (Pierce et al. 2011). This criterion is working both for
heavily obscured AGN, $N_H\lesssim10^{23}cm^{-2}$ and of being free
of host galaxy contamination.

\item $P_{5GHz}$: Goldschmidt et al. (1999) proposed a criterion
based solely on radio power, where sources with $P_{5GHz}=\log[P_{5
GHz}(W/Hz/Sr)]>24$ are considered to be RL. Given that Goldschmidt
et al. (1999) assume a Hubble parameter
H$_0$=50~km~s$^{-1}$~Mpc$^{-1}$, this criterion corresponds to
$P_{5GHz}=\log[P_{5 GHz}(W/Hz/Sr)]>23.7$ for the cosmology used in
this paper.

\end{enumerate}

The distributions of the nine radio loudness measures are shown in
Figure~\ref{rlhist}. From these plots we could only see continuous
distributions with long tail on the radio loud side and there is no
clear sign of bimodality. The size of the sample is still too small
to give statistical significant check on the bimodality of the radio
loudness measures.

\section{Radio-Loud Fraction}
\label{s:rlf}

\begin{table}
\begin{minipage}{\columnwidth}
\centering \caption{Radio-Loud Quasars by Different
Criteria\label{t:rql1}}
\begin{tabular}
{|c|c|c|c|c|} \hline criterion & N(RL) & N(RQ) & N(amb)$^{*}$ &
Fraction(RL)\\ \hline\hline
 $R_L>1$        & 18 & 367 & 22 & 4.50\%$^{+2.73\%}_{-1.52\%}$\\\hline
 $R_{i}>1$      & 16 & 364 & 27 & 3.98\%$^{+2.67\%}_{-1.38\%}$\\\hline
 $q_{24}<0$     & 13$^{a}$ & 357 & 12 & 3.43\%$^{+1.67\%}_{-1.11\%}$\\
 $q_{24,K}<0.24$& 25$^{a}$ & 307 & 50 & 7.22\%$^{+2.43\%}_{-2.70\%}$\\\hline
 $R_{uv}>1$     & 45 & 283 & 79 &11.75\%$^{+4.02\%}_{-2.60\%}$\\
 $R_{uv, D}>1.88$& 9 & 398 & 0  & 2.21\%$^{+2.09\%}_{-1.00\%}$\\\hline
 $R_X>-3$       & 10 & 397 & 0  & 2.46\%$^{+1.88\%}_{-1.23\%}$\\\hline
 $P_{5GHz}>23.7$&  8 & 399 & 0  & 1.97\%$^{+1.83\%}_{-1.03\%}$\\\hline\hline
 $R_{L,obs}>1$  & 67 & 220 &120 &19.23\%$^{+4.92\%}_{-3.58\%}$\\
 $R_{J}>1$      & 16 & 360 & 31 & 4.02\%$^{+2.71\%}_{-1.39\%}$\\\hline
 $R_{i, obs}>1$ & 38 & 300 & 69 &10.03\%$^{+3.79\%}_{-2.43\%}$\\
 $R_{K}>1$      & 10 & 392 & 5  & 2.46\%$^{+2.23\%}_{-1.04\%}$\\\hline
 $q_{24,obs}<0$ &  8 & 370 & 4$^{b}$  & 2.55\%$^{+1.49\%}_{-1.06\%}$\\\hline
 \end{tabular}\\
\end{minipage}
$^{*}$Number of ambiguous sources that with upper/lower limits in
the radio-loud region. See \S~\ref{s:rlf} for details.\\
$^{a, b}$The 2 sources (XMM ID: 320 and 5315) with VLA detection and
MIPS upper limits have upper limits on $q_{24}$ and $q_{24,obs}$.
Therefore: (a) they are already located in the RL region with upper
limits, so they are RL by the $q_{24}$ criterion; (b) they are
located in the RQ region with upper limits, so they can still be
RL, so they are ambiguous for the $q_{24,obs}$ criterion.\\
\end{table}

\begin{table*}
\begin{minipage}{145mm}
\centering \caption{Radio-Loud Quasars with Two
Criteria\label{t:rql2}}
\begin{tabular}
{c|c|c|c|cc|c} \hline criterion & N(RL) & N(RQ) &
\multicolumn{3}{|c|}{N(amb)$^{*}$} & Fraction(RL)\\ \hline
 & & & 1 & \multicolumn{2}{|c|}{2} & \\ \hline\hline
 $R_L>1$       &                          &                            & 10 & 8  &  &  \\
 $q_{24}<0$    & \raisebox{1.3ex}[0pt]{8} & \raisebox{1.3ex}[0pt]{339} &  5$^{a}$ & 8 &\raisebox{1.3ex}[0pt]{4}& \raisebox{1.3ex}[0pt]{2.1\%}\\
 \hline
 $R_L>1$       &                          &                            & 0(10) & 0(22)  &  &  \\
 $R_{uv}>1(1.88)^{**}$& \raisebox{1.3ex}[0pt]{18(8)}& \raisebox{1.3ex}[0pt]{283(366)} & 27(1)& 57(0) &\raisebox{1.3ex}[0pt]{22(0)}& \raisebox{1.3ex}[0pt]{4.4\%(2.0\%)}\\
 \hline
 $R_L>1$       &                          &                            & 4 & 5  &  &  \\
 $R_i>1$       & \raisebox{1.3ex}[0pt]{14}& \raisebox{1.3ex}[0pt]{355} & 2 & 10 &\raisebox{1.3ex}[0pt]{17}& \raisebox{1.3ex}[0pt]{3.4\%}\\
 \hline
 $R_L>1$       &                          &                            & 11 & 22  &  &  \\
 $R_X>-3$      & \raisebox{1.3ex}[0pt]{7} & \raisebox{1.3ex}[0pt]{364} & 3 & 0 &\raisebox{1.3ex}[0pt]{0}& \raisebox{1.3ex}[0pt]{1.7\%}\\
 \hline
 $R_L>1$       &                          &                            & 12 & 22 &  & \\
 $P_{5GHz}>23.7$&\raisebox{1.3ex}[0pt]{6} & \raisebox{1.3ex}[0pt]{365} & 2  & 0 & \raisebox{1.3ex}[0pt]{0} & \raisebox{1.3ex}[0pt]{1.5\%}\\
 \hline
 $q_{24}<0$    &                          &                            & 0(6$^{a}$) & 5(12) &  & \\
 $R_{uv}>1(1.88)^{**}$&\raisebox{1.3ex}[0pt]{13$^{b}$(7)}&\raisebox{1.3ex}[0pt]{270(355)}& 32(2)  & 55(0) & \raisebox{1.3ex}[0pt]{7(0)} &\raisebox{1.3ex}[0pt]{3.4\%(1.8\%)}\\
 \hline
 $q_{24}<0$    &                          &                            & 5$^{a}$& 7 &  & \\
 $R_{i}>1$     & \raisebox{1.3ex}[0pt]{8} & \raisebox{1.3ex}[0pt]{342} & 8  & 7 & \raisebox{1.3ex}[0pt]{5} &\raisebox{1.3ex}[0pt]{2.1\%}\\
 \hline
 $q_{24}<0$    &                          &                            & 6$^{a}$& 12 &  & \\
 $R_{X}>-3$    & \raisebox{1.3ex}[0pt]{7} & \raisebox{1.3ex}[0pt]{354} & 3  & 0 & \raisebox{1.3ex}[0pt]{0} &\raisebox{1.3ex}[0pt]{1.8\%}\\
 \hline
 $q_{24}<0$    &                          &                            & 7$^{a}$ & 12 &  & \\
 $P_{5GHz}>23.7$&\raisebox{1.3ex}[0pt]{6} & \raisebox{1.3ex}[0pt]{355} & 2  & 0 & \raisebox{1.3ex}[0pt]{0} &\raisebox{1.3ex}[0pt]{1.6\%}\\
 \hline
 $R_{uv}>1(1.88)^{**}$&                          &                            & 30(2) & 57(0)  &  &  \\
 $R_i>1$       & \raisebox{1.3ex}[0pt]{15(7)}& \raisebox{1.3ex}[0pt]{277(362)} & 1(9)  & 5(27) &\raisebox{1.3ex}[0pt]{22(0)}& \raisebox{1.3ex}[0pt]{3.7\%(1.7\%)}\\
 \hline
 $R_{uv}>1(1.88)^{**}$&                          &                            & 37(3) & 79(0) &  &  \\
 $R_X>-3$      & \raisebox{1.3ex}[0pt]{8(6)} & \raisebox{1.3ex}[0pt]{281(394)} & 2(4) & 0(0) &\raisebox{1.3ex}[0pt]{0(0)}& \raisebox{1.3ex}[0pt]{2.0\%(1.5\%)}\\
 \hline
 $R_{uv}>1(1.88)^{**}$&                          &                            & 39(4) & 79(0) &  & \\
 $P_{5GHz}>23.7$& \raisebox{1.3ex}[0pt]{6(5)} & \raisebox{1.3ex}[0pt]{281(395)} & 2(3) & 0(0) & \raisebox{1.3ex}[0pt]{0(0)} & \raisebox{1.3ex}[0pt]{1.5\%(1.2\%)}\\
 \hline
 $R_i>1$       &                          &                            & 7 & 27  &  &  \\
 $R_X>-3$      & \raisebox{1.3ex}[0pt]{9} & \raisebox{1.3ex}[0pt]{363} & 1 & 0 &\raisebox{1.3ex}[0pt]{0}& \raisebox{1.3ex}[0pt]{2.2\%}\\
 \hline
 $R_i>1$       &                          &                            & 10 & 27 &  & \\
 $P_{5GHz}>23.7$&\raisebox{1.3ex}[0pt]{6} & \raisebox{1.3ex}[0pt]{362} & 2  & 0 & \raisebox{1.3ex}[0pt]{0} & \raisebox{1.3ex}[0pt]{1.5\%}\\
 \hline
 $R_X>-3$      &                          &                            & 3 & 0 &  & \\
 $P_{5GHz}>23.7$& \raisebox{1.3ex}[0pt]{7}& \raisebox{1.3ex}[0pt]{396} & 1 & 0 & \raisebox{1.3ex}[0pt]{0} & \raisebox{1.3ex}[0pt]{1.7\%}\\
 \hline\hline
 $R_{L,obs}>1$ &                          &                            & 57 & 95  &  & \\
 $q_{24,obs}<0$& \raisebox{1.3ex}[0pt]{8} & \raisebox{1.3ex}[0pt]{218} &  0 & 2$^{c}$  &\raisebox{1.3ex}[0pt]{2}& \raisebox{1.3ex}[0pt]{2.1\%}\\
 \hline
 $R_{i,obs}>1$ &                          &                            & 28 & 49  &  & \\
 $q_{24,obs}<0$& \raisebox{1.3ex}[0pt]{8} & \raisebox{1.3ex}[0pt]{293} &  0 & 2$^{c}$  &\raisebox{1.3ex}[0pt]{2}& \raisebox{1.3ex}[0pt]{2.1\%}\\
 \hline
 $R_{L,obs}>1$ &                          &                            & 29 & 51  &  & \\
 $R_{i,obs}<0$& \raisebox{1.3ex}[0pt]{38} & \raisebox{1.3ex}[0pt]{220} &  0 & 0  &\raisebox{1.3ex}[0pt]{69}& \raisebox{1.3ex}[0pt]{9.3\%}\\
 \hline
 \end{tabular}\\
$^{*}$Number of ambiguous sources: 1) they are RL with one
criterion, but RQ with the other; 2) the upper/lower limits locate
in the radio-loud region for one or both criteria, that they could
be RQ. See \S~\ref{s:rlf} for details.\\
$^{**}$ The numbers in the parenthesis are for the $R_{uv,D}$ criterion.\\
$^{a, b, c}$The 2 sources (XMM ID: 320 and 5315) with VLA detection
and MIPS upper limits have upper limits on $q_{24}$ and
$q_{24,obs}$. Therefore: (a) they are already located in the RL
region with upper limits, so they are RL by the $q_{24}$ criterion.
But they are not RL in the other criterion, so they are still
ambiguous sources; (b) they are already located in the RL region
with upper limits, so they are RL by the $q_{24}$ criterion. And
they are RL by the $R_{uv}$ criterion, so they are not ambiguous
sources; (c) they are located in the RQ region with the upper
limits, so they can still be RL, so they are ambiguous for the
$q_{24,obs}$ criterion.
\end{minipage}
\end{table*}

\subsection{Single Criterion}

The numbers and fractions of RL quasars in the sample using the nine
different RL selection criteria are summarized in
Table~\ref{t:rql1}. The RL-fraction is calculated (a) using
Kaplan-Meier product limit estimator (Kaplan \& Meier 1958) if the
sample is singly-censored or (b) following the iterative procedure
in Schmitt (1985) if both upper and lower limits on the loudness
diagnostic are present ($q_{24}$ and $q_{24, obs}$). We also list
the number of ambiguous sources in the table, which are those with
upper/lower limits lying in the RL region.

The fraction of radio-loud quasars spans a wide range from $\sim2\%$
to $\sim20\%$.

\begin{figure}
\epsfig{file=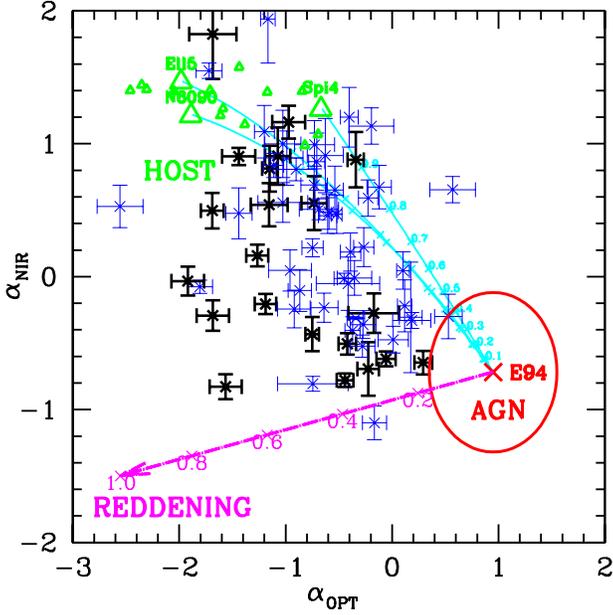, angle=0,width=\linewidth}
\caption{Mixing diagram (Hao et al. 2012) for RL quasars according
to the $R_{uv}$ criterion, but classified as RQ by all other
selections. Black points are used for quasars with radio detections,
blue points for ``ambiguous'' quasars (upper limit on $R_{uv}$
exceeding threshold $R_{uv}$=1). The mixing diagram is the plot of
the SED slopes $\alpha_{NIR}$ ($3\mu m$ to $1\mu m$) versus
$\alpha_{OPT}$ ($1\mu m$ to $0.3\mu m$). The E94 RQ mean SED is
shown as the red cross. The red circle shows the dispersion of E94.
The green triangles show galaxy templates from the SWIRE template
library (Polletta et al. 2007). The purple arrow represents the
reddening vector for the E94 RQ mean SED. The cyan lines connect the
galaxy templates and the E94 mean SED are mixing curves showing
templates with different galaxy fractions. \label{slpplotrql}}
\end{figure}

For the criteria defined in the rest-frame, $R_{uv}$ classifies the
largest number ($\sim12\%$) of COSMOS AGNs as RL, where most of
these quasars are RQ by all other criteria defined in the rest
frame. The SEDs of the quasars classified as RL by $R_{uv}$ but not
the other criteria generally do not show a `big blue bump' feature
in their SEDs that is characteristic of unobscured quasars (Paper
I). We plot the sources which have either a direct measurement or
upper limit of $R_{uv}$ that exceeds the selection threshold for
radio loudness ($R_{uv}=1$), but which are RQ according to all
alternative definitions defined in the rest-frame, in the Hao et al.
(2013) mixing diagram (Figure~\ref{slpplotrql}). These quasars are
mainly located in the high reddening ($E(B-V)>0.2$) or high galaxy
fraction ($f_g>0.4$) regions and well away from the E94 SED region
(red circle in Figure~\ref{slpplotrql}). This suggests that they are
mostly reddened or galaxy dominated sources and their apparent
radio-loudness by $R_{uv}$ is due to these contaminating factors.
Note that if we change the RQ and RL dividing line, i.e. if we use
the $R_{uv, D}$ criterion instead, the number of ambiguous sources
would be much lower and the radio-loud fraction will drop to
$\sim2\%$. This criterion is more appropriate for X-ray selected
samples.

Therefore, for the criteria defined in the rest-frame, the RL
fraction is 2.0\%--4.5\% using any single criterion (for $R_{uv}$
considering only the $R_{uv,D}$). This is small compared to the
$\sim$10\% seen in typical optically selected AGN samples (e.g.
Peterson et al. 1997). To reach 10\% would require an additional
22--32 AGNs to be classified as RL. The $P_{5GHz}$ ($\sim$2.0\%) is
the most restrictive definition of radio-loudness, while $R_L$
($\sim$4.5\%) gives somewhat larger RL samples.

For the criteria defined in the observed frame, we note that no
k-correction is included, so for quasars at different redshift, the
ratio is actually calculated at a different frequency. The radio
loudness definition with radio-to-optical ratio all yield a high
radio-loud fraction ($>10\%$) in contrast to all the other criteria.
This is because the quasars in the sample have a large redshift
range ($0.1\leq z\leq4.3$). If we consider that most of the quasars
in the XMM-COSMOS sample are at redshift 1--2, the rest-frame B
band, i band and 5GHz would be at J band, K band and 1.4GHz in the
observed frame, then we can define $R_{J}$ and $R_{K}$ instead of
$R_{L, obs}$ and $R_{i,obs}$, the radio-loud fraction reduce to
$\sim 4\%$ and $\sim 2.5\%$ respectively. Although $q_{24,obs}$ also
does not include k-correction, it still yields a low radio-loud
fraction (2.55\%).

\subsection{Pairs of Criteria}
We also consider 18 pairs of RL criteria (Table~\ref{t:rql2}). To
avoid confusion with k-correction effects, criteria in the observed
frame are not compared with criteria in the rest-frame. Thus, these
18 pairs include all the possible combinations of criteria pairs.
The RL fraction is determined by the number of objects that lie in
the RL-selection region for both measures of radio-loudness divided
by the total number of sources in the sample (407; when $q_{24}$ or
$q_{24,obs}$ included, it is 382). The ``ambiguous'' sources
include: 1. sources identified as RL by one criterion (detections
only) but RQ with the other; 2. sources with only upper/lower limits
which lie in the RL region.

If we require any two criteria to agree, then the RL fractions are
even smaller (1.5\%--4.4\%, Table~\ref{t:rql2}) except for
$R_{L,obs}$ and $R_{i,obs}$. For about 3/4 of all possible
combinations, the RL fraction are smaller or around 2\%.

\subsection{Solidly Radio Loud Quasars}

\begin{figure*}
\includegraphics[angle=0,width=0.32\textwidth]{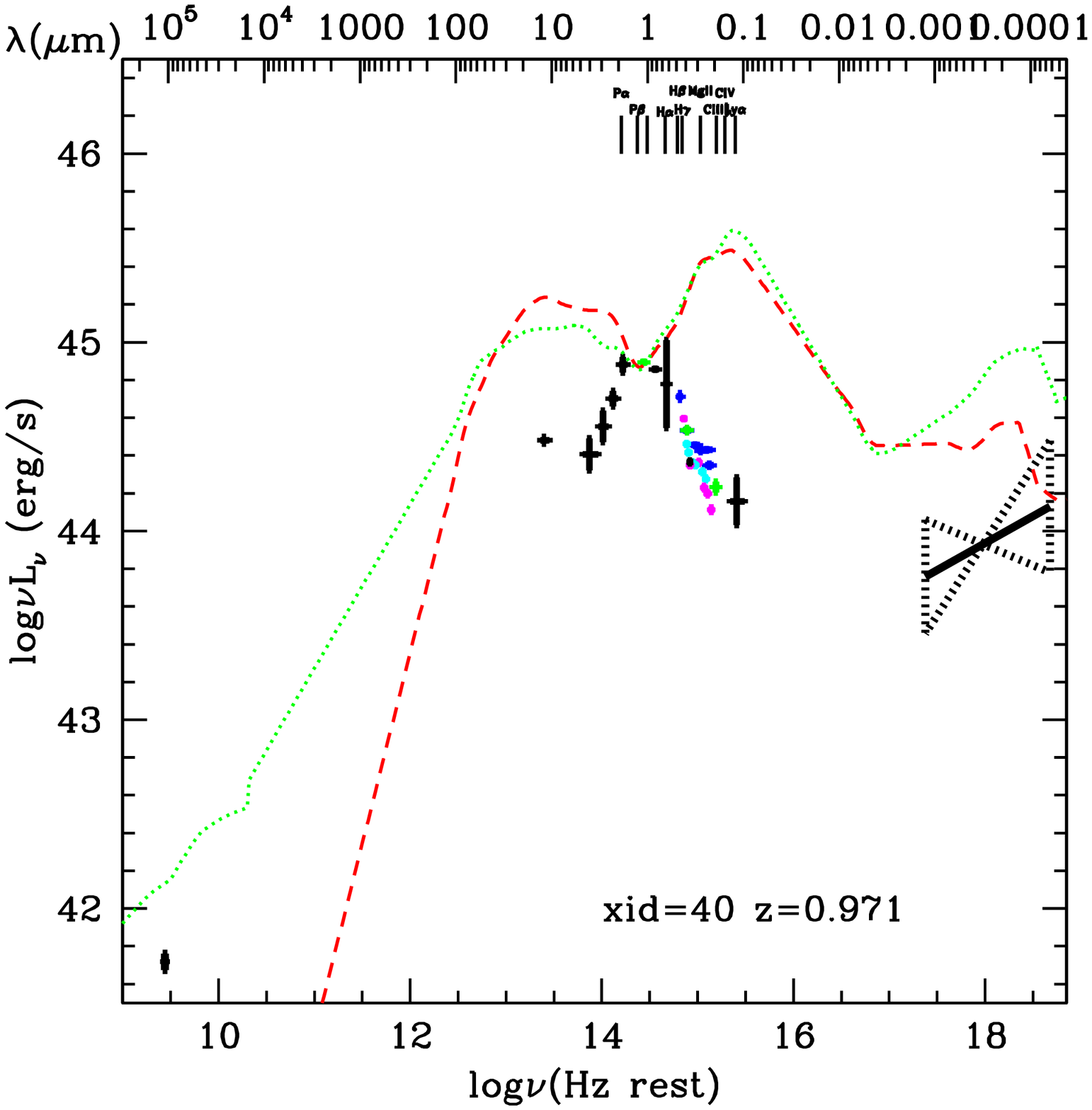}
\includegraphics[angle=0,width=0.32\textwidth]{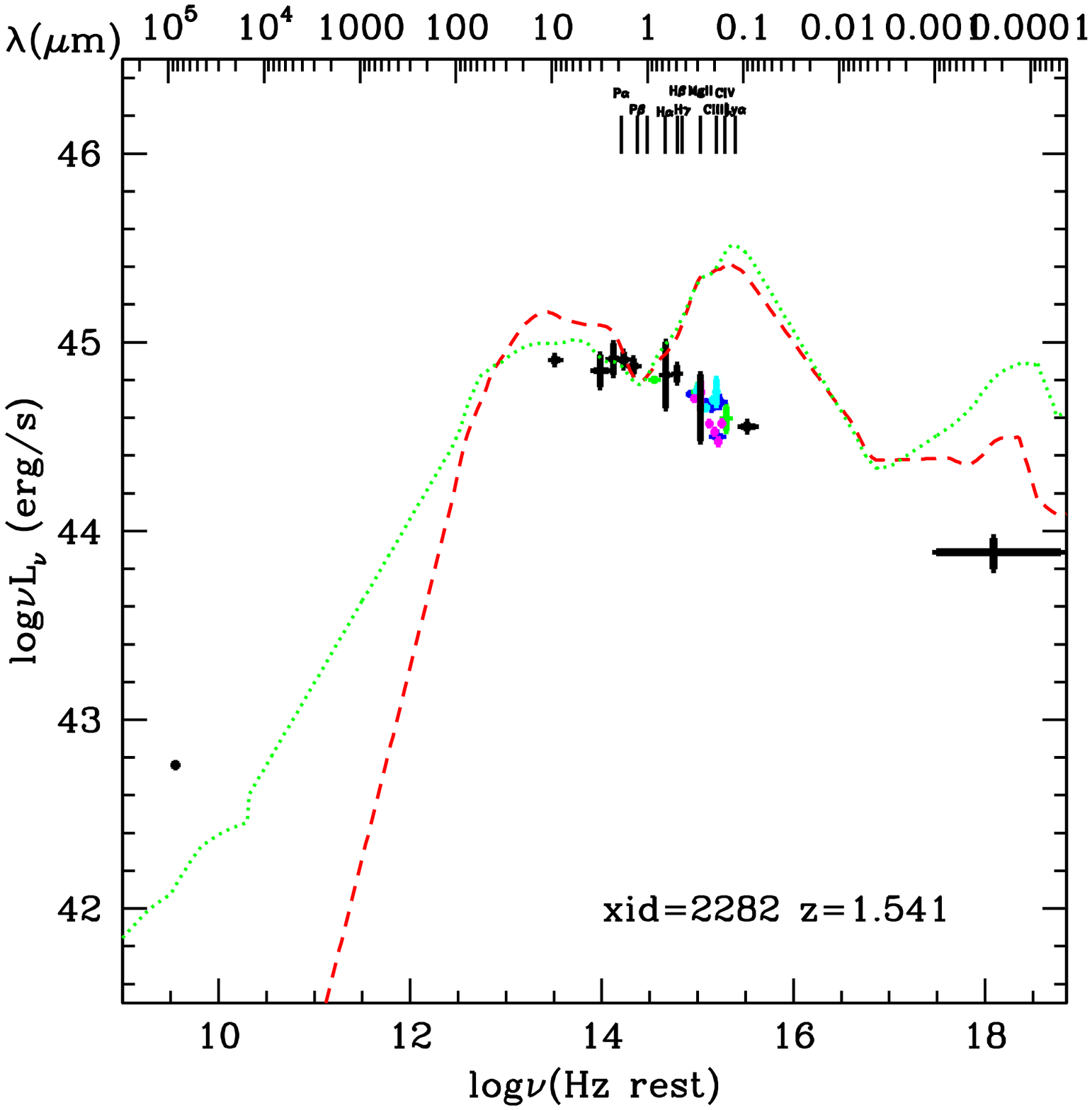}
\includegraphics[angle=0,width=0.32\textwidth]{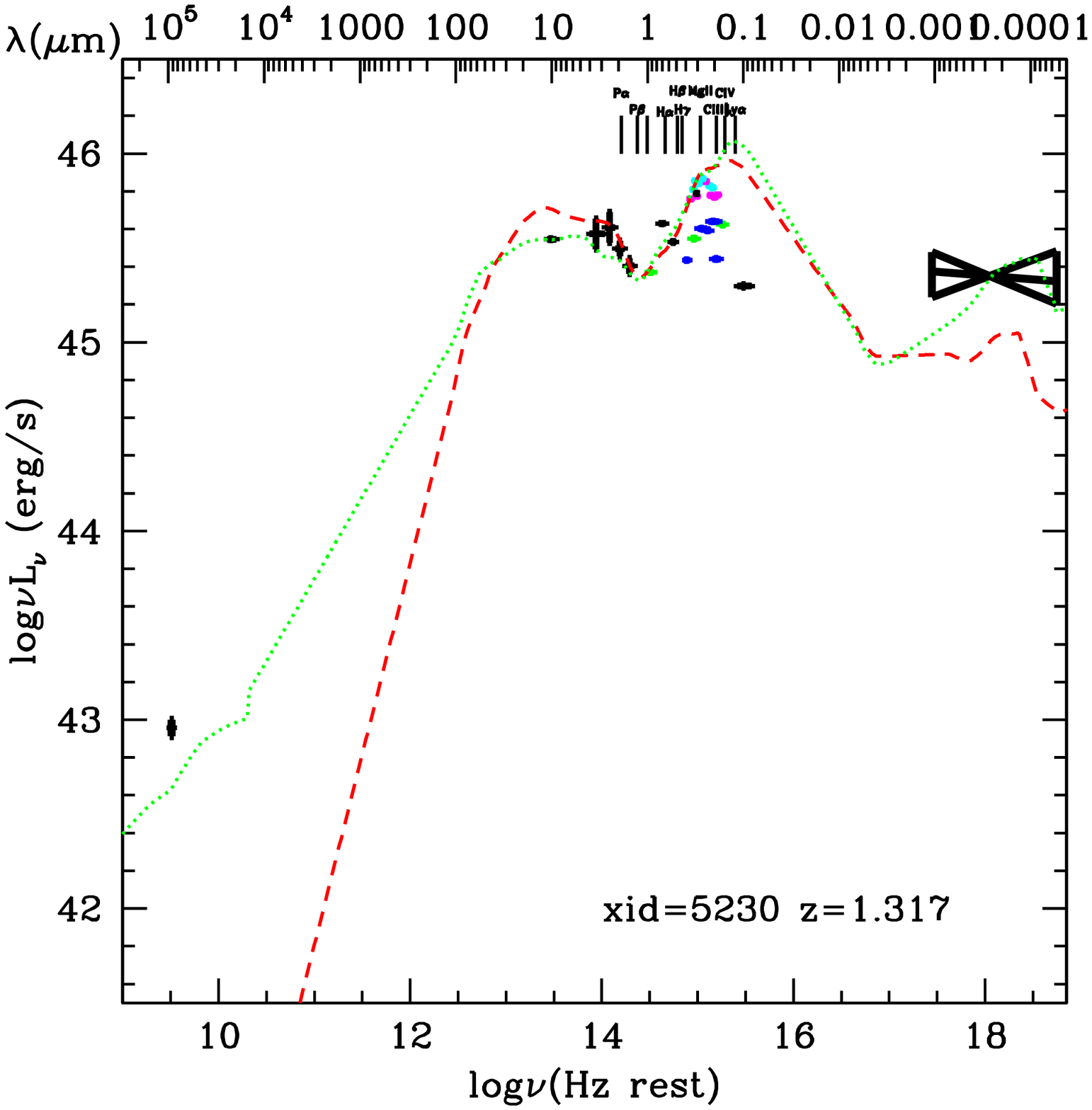}
\includegraphics[angle=0,width=0.32\textwidth]{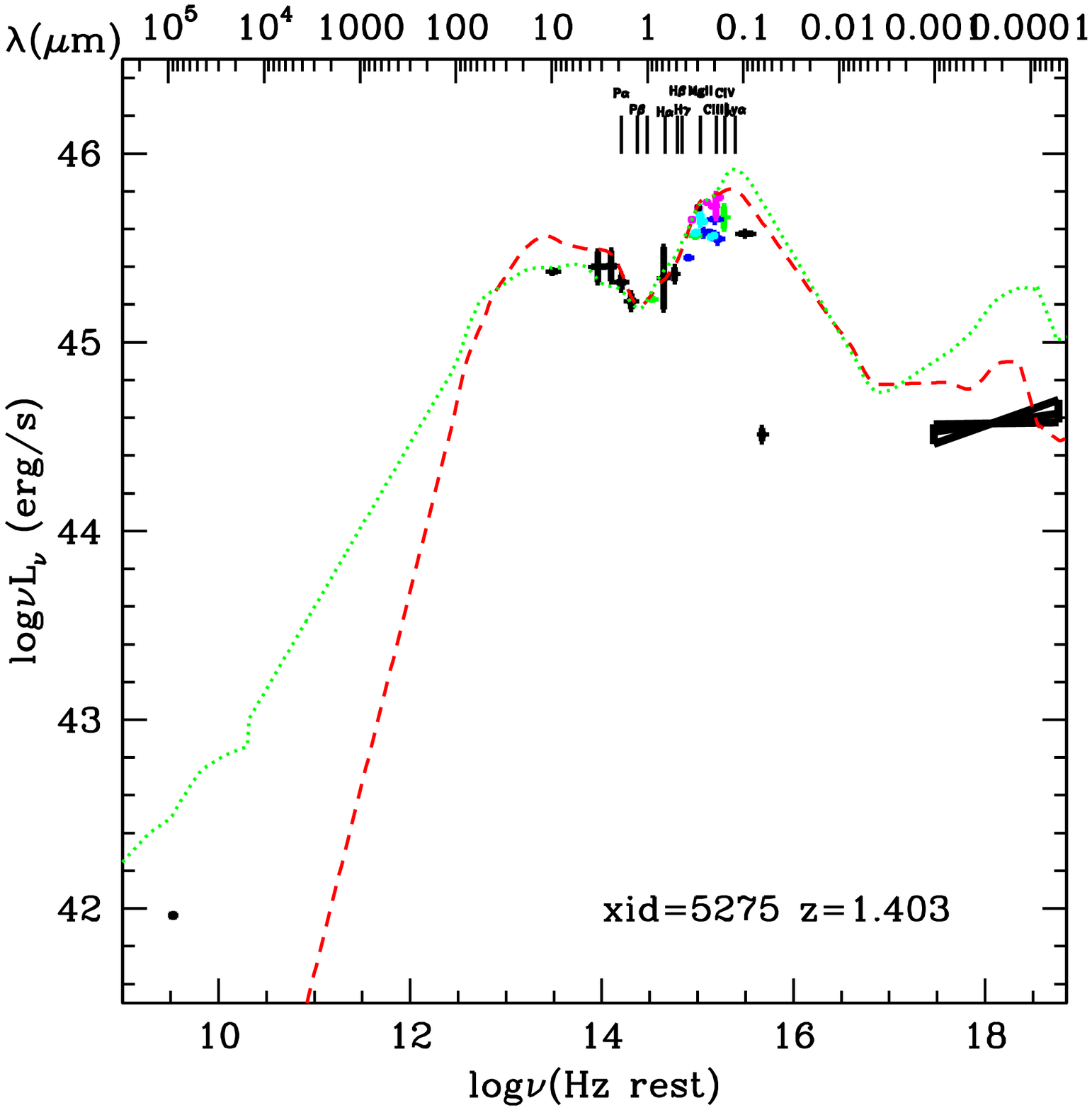}
\includegraphics[angle=0,width=0.32\textwidth]{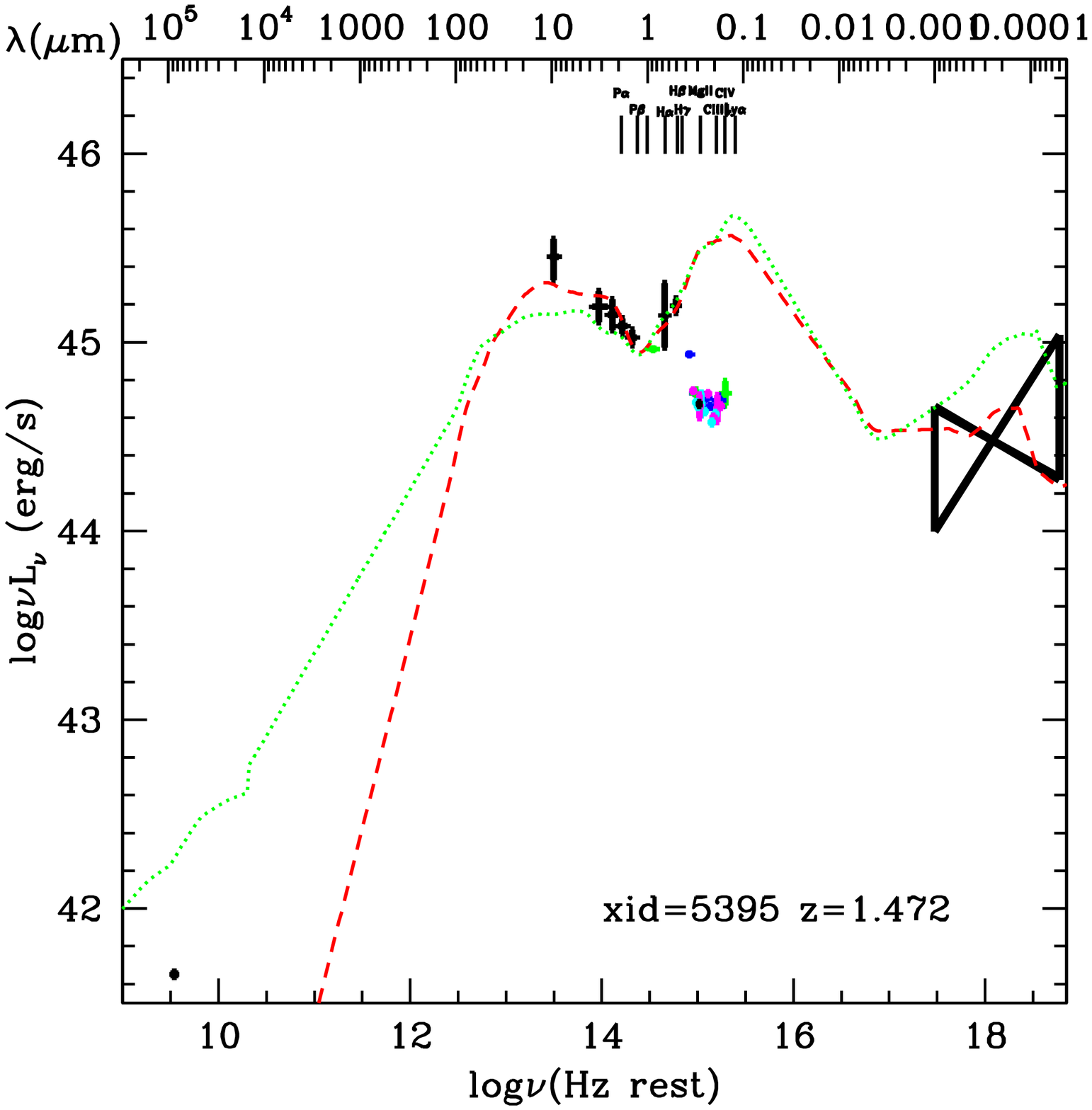}
\includegraphics[angle=0,width=0.32\textwidth]{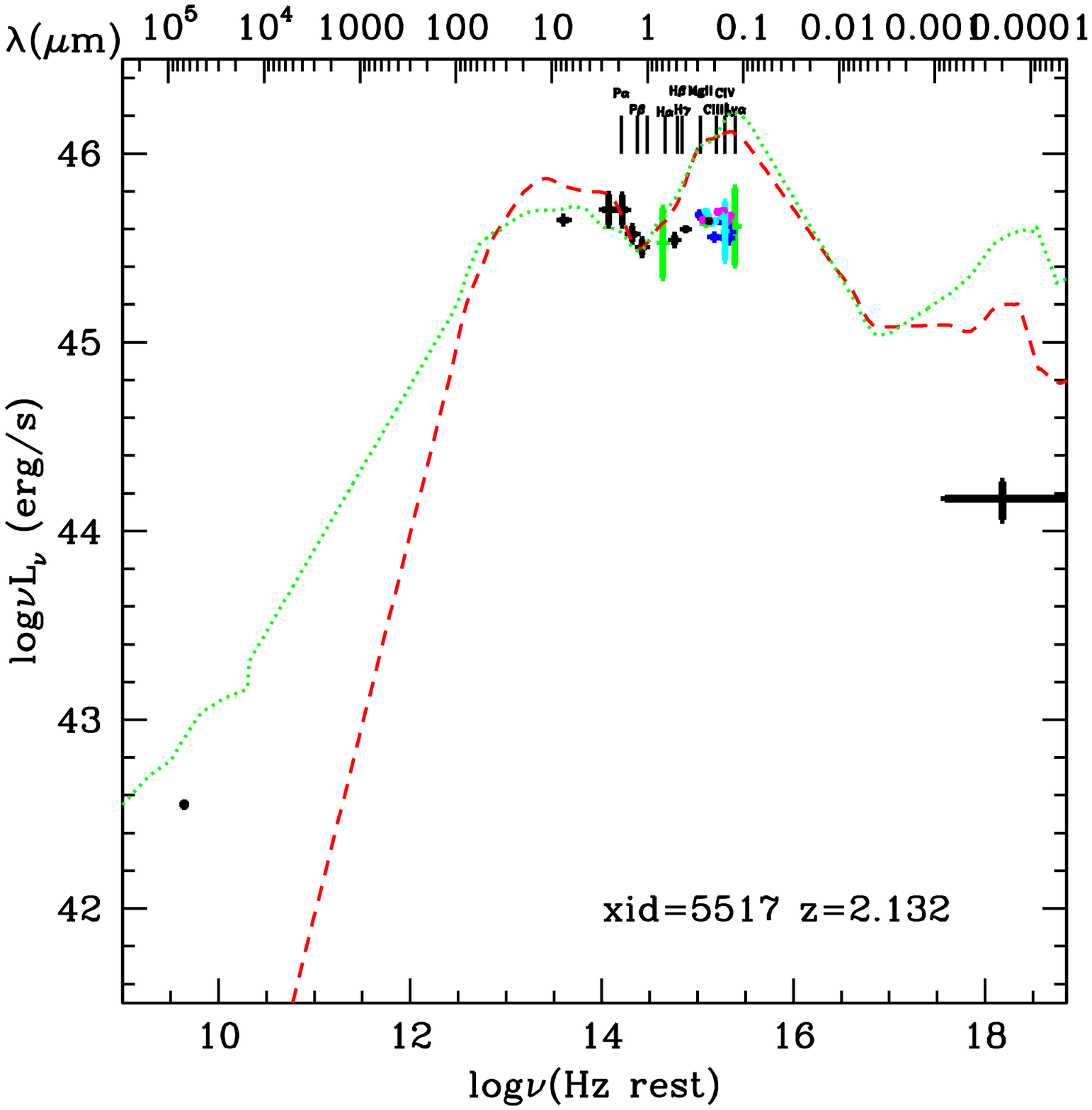}
\caption{The spectral energy distribution (SED) of the six quasars
which are radio-loud in all criteria.  The red dashed line is the
E94 RQ mean SED. The green dotted line is the E94 RL mean SED. The
data points in the SED are color-coded as in Elvis et al. (2012).
From low to high frequency, the black data points are: 1.4GHz,
24$~\mu$m, 8$~\mu$m, 5.7$~\mu$m, 4.5$~\mu$m, 3.6$~\mu$m, K-band,
H-band, J-band, NUV, FUV and 2keV. The blue data points are the
Subaru broad bands ($\rm{B_J}$, g, r, i, z). The green data points
are the (CFHT) K-band, and the (CFHT) u band and i band. The purple
data points are the 6 Subaru intermediate bands for season 1 (2006)
(IA427, IA464, IA505, IA574, IA709, IA827).  The cyan data points
are the 5 Subaru intermediate bands for season 2 (2007) (IA484,
IA527, IA624, IA679, IA738, IA767). \label{rlsed}}
\end{figure*}

\begin{table}
\begin{minipage}{\columnwidth}
\centering \caption{Radio-Loud Quasar Properties\label{t:rlprop}}
\begin{tabular}
{|c|c|c|c|c|} \hline XID & z & $\log(L_{bol})^{*}$ &
$\log (M_{BH}/M_{\bigodot})^{**}$ & $\lambda_{Edd}^{***}$\\
\hline\hline

 40   & 0.971 & 45.24 & $\cdots$ & $\cdots$ \\
 2282 & 1.541 & 45.38 & $\cdots$ & $\cdots$ \\
 5230 & 1.317 & 46.44 & 8.21   & 1.337  \\
 5257 & 1.403 & 46.07 & 9.06   & 0.080  \\
 5395 & 1.472 & 45.68 & $\cdots$ & $\cdots$ \\
 5517 & 2.132 & 46.24 & 8.70   & 0.277  \\
 \hline
\end{tabular}
\end{minipage}
$^{*}$$\log(L_{bol})$ is calculated by integrating the rest-frame
SED from $1~\mu m$ to 40~keV.\\
$^{**}$ The black hole mass estimates are from Trump
et al. (2009b) and Merloni et al. (2010).\\
$^{***}$Eddington ratio
$$\lambda_{Edd}=\frac{L_{bol}}{L_{Edd}}=\frac{L_{bol}}{\frac{4\pi
Gcm_{p}}{\sigma_e}M_{BH}}
=\frac{L_{bol}}{1.26\times10^{38}(M_{BH}/M_{\odot})}$$
\end{table}

Finally, if we require all the above criteria to be satisfied
simultaneously, just six sources remain\footnotemark. The according
RL fraction is thus 1.5\% (6 out of 407), only marginally lower than
that obtained for a classification with the $P_{5GHz}$ criterion
alone. We plot the SEDs of these quasars in Figure~\ref{rlsed}. The
general properties of these quasars are listed in
Table~\ref{t:rlprop}. For the three of the six, black hole estimates
are available and all exceed $10^8 M_{\bigodot}$.
\footnotetext{XID=40, 2282, 5230, 5275, 5517, 5395. Note if we use
the $R_{uv, D}$ criterion, the quasar XID=5275 is classified as RQ.
}

\subsection{Inter-comparison of Criteria}
\begin{figure*}
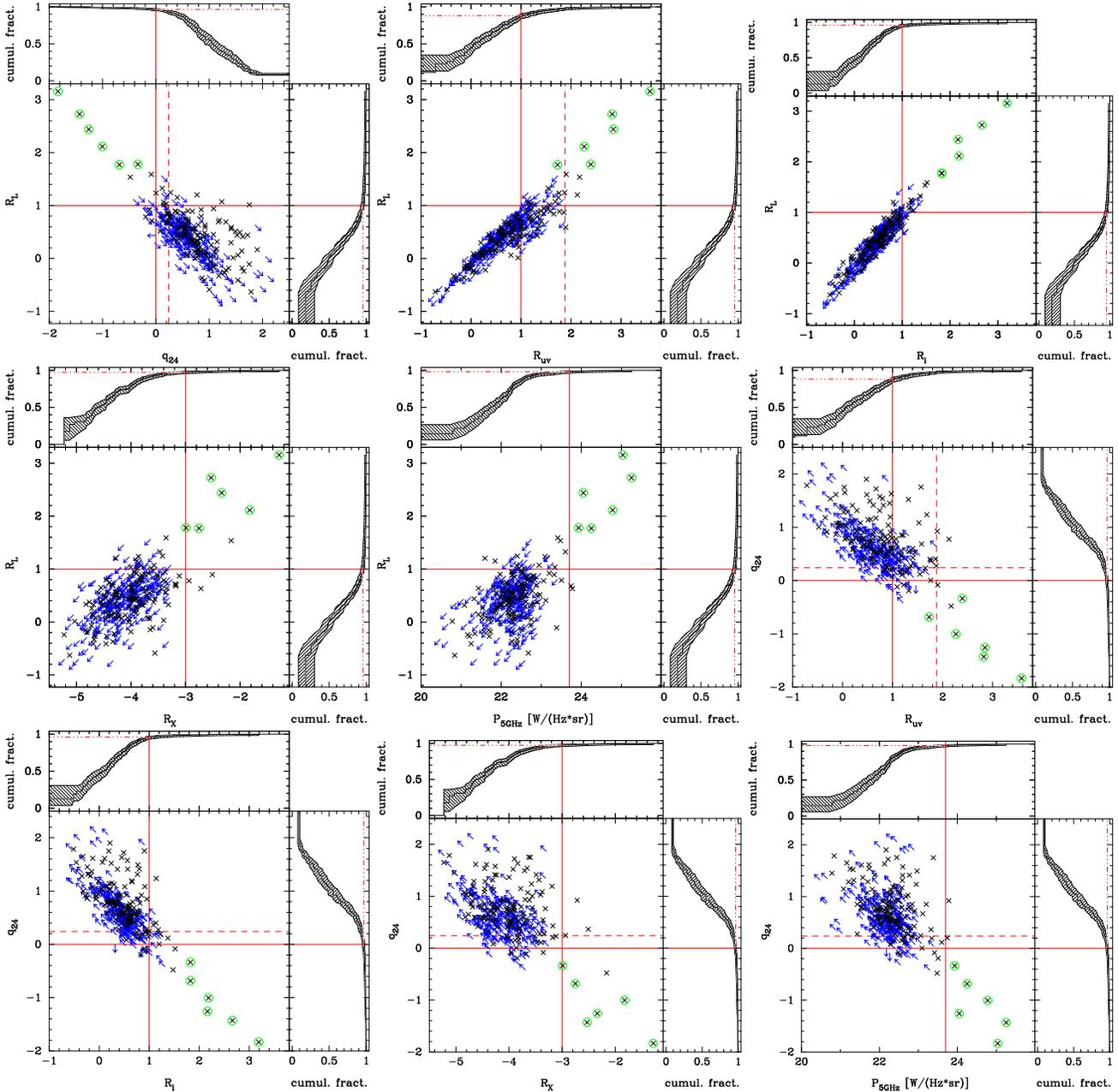

\includegraphics[angle=0,width=0.32\textwidth]{RLvsq24_v1.ps}
\includegraphics[angle=0,width=0.32\textwidth]{RLvsRuv_v1.ps}
\includegraphics[angle=0,width=0.32\textwidth]{RLvsRLi_v1.ps}
\includegraphics[angle=0,width=0.32\textwidth]{RLvsRx_v1.ps}
\includegraphics[angle=0,width=0.32\textwidth]{RLvsP5_v1.ps}
\includegraphics[angle=0,width=0.32\textwidth]{q24vsRuv_v1.ps}
\includegraphics[angle=0,width=0.32\textwidth]{q24vsRLi_v1.ps}
\includegraphics[angle=0,width=0.32\textwidth]{q24vsRx_v1.ps}
\includegraphics[angle=0,width=0.32\textwidth]{q24vsP5_v1.ps}
\caption{Radio-loudness measures: $R_L$, $q_{24}$, $R_{uv}$, $R_i$,
$R_X$, and $P_{5GHz}$. Black crosses = radio detections; blue arrows
= upper/lower limits; green circles = RL in all criteria. Solid
lines = limits assumed for the radio-loudness definition. Dashed
line perpendicular to $q_{24}$ axis = Kuraszkiewicz et al. (in
preparation) adjusted $q_{24}$ criterion. Dashed line perpendicular
to $R_{uv}$ axis = Della Ceca et al. (1994) adjusted $R_{uv}$
criterion. Along the upper/right-hand edge we show the cumulative
distribution functions which include a correction for upper/lower
limits. Dash-dotted lines indicate the RQ fraction. \label{rlfig1}}
\end{figure*}

\begin{figure*}
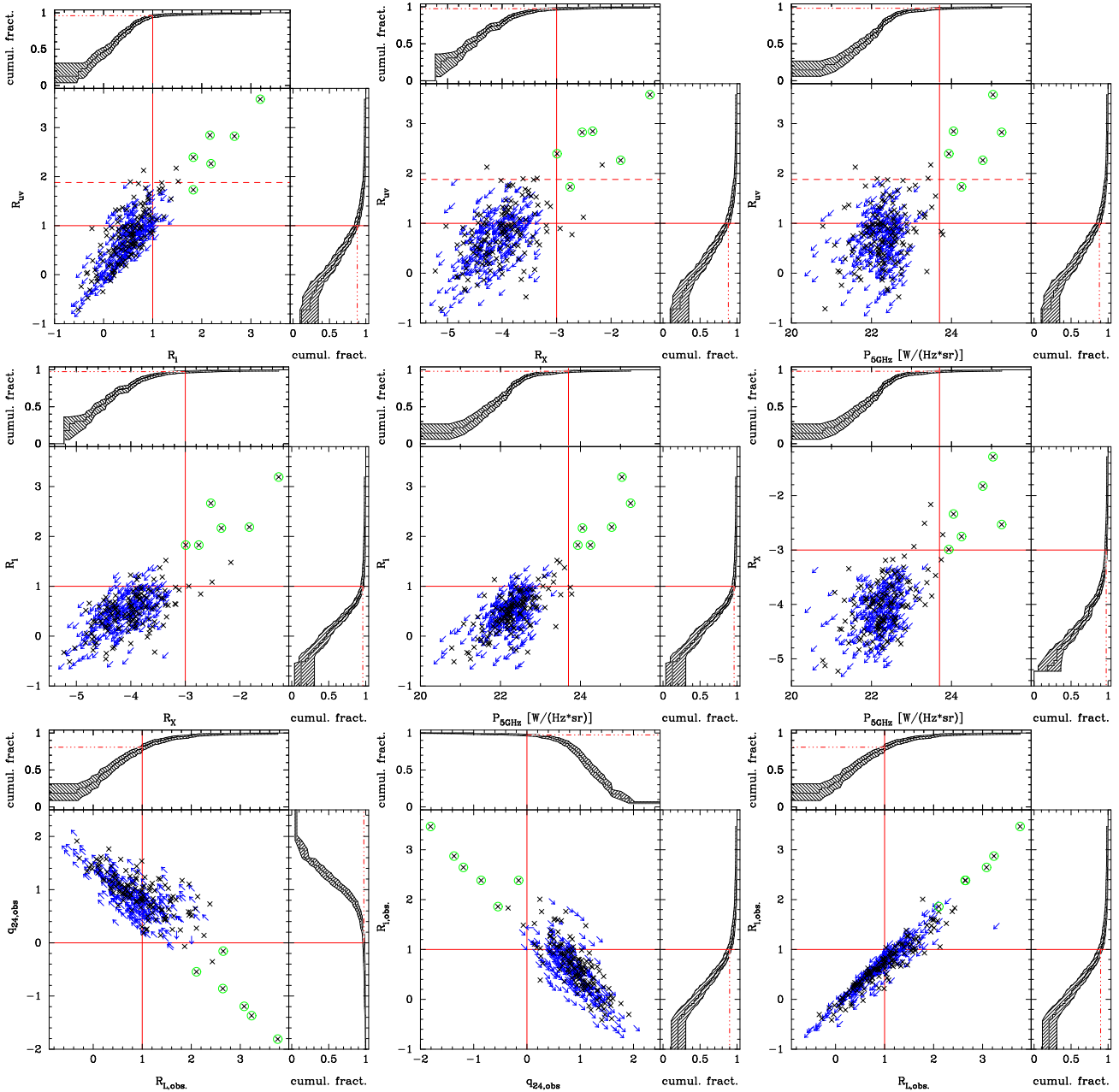

\includegraphics[angle=0,width=0.32\textwidth]{RuvvsRLi_v1.ps}
\includegraphics[angle=0,width=0.32\textwidth]{RuvvsRx_v1.ps}
\includegraphics[angle=0,width=0.32\textwidth]{RuvvsP5_v1.ps}
\includegraphics[angle=0,width=0.32\textwidth]{RivsRx_v1.ps}
\includegraphics[angle=0,width=0.32\textwidth]{RivsP5_v1.ps}
\includegraphics[angle=0,width=0.32\textwidth]{RxvsP5_v1.ps}
\includegraphics[angle=0,width=0.32\textwidth]{q24obsvsRLo_v1.ps}
\includegraphics[angle=0,width=0.32\textwidth]{RLiobsvsq24obs_v1.ps}
\includegraphics[angle=0,width=0.32\textwidth]{RLiobsvsRLo_v1.ps}
\caption{Radio-loudness measures: $R_{uv}$, $R_i$, $R_X$,
$P_{5GHz}$, $R_{L, obs}$, $q_{24, obs}$ and $R_{i, obs}$. Black
crosses = radio detections; blue arrows = upper/lower limits; green
circles = RL in all criteria. Solid lines = limits assumed for the
radio-loudness definition. Dashed line = Della Ceca et al. (1994)
adjusted $R_{uv}$ criterion. Along the upper/right-hand edge we show
the cumulative distribution functions which include a correction for
upper/lower limits. Dash-dotted lines indicate the RQ fraction.
\label{rlfig2}}
\end{figure*}

Figure~\ref{rlfig1} \& \ref{rlfig2} show the distributions of the
sources with respect to the different RL diagnostics to illustrate
the RL fraction and compare the agreement among the above radio
loudness definitions. On the side of each axis, we also plot the
cumulative fraction of the sources as function of the corresponding
radio-loudness measurements using the survival analysis methods
(Kaplan \& Meier 1958; Schmitt 1985) which we previously employed to
compute the RL fractions. We marked the 6 quasars in the sample that
are classified as RL according to all nine criteria with green
circles. Their SEDs are plotted in Figure~\ref{rlsed}.

For the criteria defined in the rest-frame, the radio-loudness
defined by the ratios of radio to optical luminosity ($R_L$,
$R_{uv}$ and $R_i$) correlate well with each other
(Figure~\ref{rlfig1} \& \ref{rlfig2}). The $R_X$ and $P_{5GHz}$
criteria agree the best with the fewest `ambiguous' quasars.

\section{Discussion}

\subsection{Low Radio-Loud Fraction}
We have found a low RL fraction of 1.5\%--4.5\% (using six different
criteria defined in the rest-frame) in the XMM-COSMOS type 1 AGN
sample, as compared with about 10\% in optically selected samples.
For example, in the BQS, Kellermann et al. (1989), find 15\% using
the $R_L$ criterion; in LBQS, Hooper et al. (1996) find 9\% using
the rest-frame 8.4 GHz luminosity $\log L_{8.4}>25$ and 9\% using
the flux ratio between the rest-frame 8.4 GHz and B band
$R_{8.4}>1$. The difference between XC407 and these other RL
fractions is significant at a confidence level of $>99\%$ and is
observed for all RL criteria. In compiling these numbers, we use the
criterion $R_{uv,D}$ instead of $R_{uv}$, as it is subject to
reddening and host contamination issues (\S \ref{s:rlf}).

Similarly low RL fractions have been reported or inferred in a few
other samples, all of which include infrared selection. For example,
Richards et al. (2006) reported only 8 RL quasars among a
Spitzer-Sloan Digital Sky Survey (SDSS) quasar sample of 259
sources, giving a similarly small RL fraction of 3\%, using the
criterion of radio luminosity $L_{rad}>10^{33}~erg~s^{-1}~Hz^{-1}$,
that is $\log[L_{rad}(W/Hz)]>26$, which is stricter than the typical
$P_{5GHz}$ criterion we applied in the present analysis.

Donley et al. (2007) found a RL fraction of 3\% when they applied
the $q_{24}$ criterion to a sample of 62 X-ray selected power-law
AGNs in the {\em Chandra} Deep Field North (CDFN) whose Spitzer IRAC
SEDs exhibit the characteristic power-law emission expected for
luminous AGNs. Donley et al. (2012) attribute this low RL fraction
to the fact that IRAC color-color selection is biased against
sources with particularly bright hosts and radio-loud AGN tend to be
hosted by bright elliptical galaxies.

Of the criterion defined in the observed frame, the criteria
$R_{L,obs}$ and $R_{i,obs}$ yield a high radio-loud fraction
($>10\%$). This number is comparable to previous studies. For
example, in the Sloan Digital Sky Survey --DR2/FIRST, Ivezi\'{c} et
al. (2002) find 8\%$\pm$1\% using the flux ratio between the 1.4 GHz
and i band in the observed frame $R_{i,obs}>1$. However, we note
that no k-correction is included in these two criteria and the
XMM-COSMOS sample has a large redshift range. Considering most of
the quasars in the XMM-COSMOS sample are at redshift 1--2, we define
$R_{J}$ and $R_{K}$ instead to ensure a more meaningful comparison
with the observed frame criteria applied to low redshift samples.
The radio-loud fraction then drops to about $4\%$ and $2.5\%$,
respectively. Even though $q_{24, obs}$ also involves no
k-correction, the corresponding RL fraction is nevertheless low
(2.55\%). Therefore, for samples with large redshift range, the
criteria defined in the observed frame not including k-correction
could give different radio-loud fraction compared to other criteria
defined in the rest-frame.

X-ray samples often select more obscured or host dominated AGNs than
optically selected samples (Hao et al. 2014, Kuraszkiewicz et al.
2003), because X-ray emission is ubiquitous in AGNs which makes
X-ray surveys the most complete census of AGNs of any single band
(Risaliti \& Elvis 2004). Reddening will increase the RL fraction,
as we confirm in the present XMM-COSMOS sample based on the $R_{uv}$
criterion. A large host galaxy contribution could artificially boost
the apparent optical flux, thereby fortuitously reducing the RL
fraction for most of these criteria (e.g. $R_L$), while the
combination effect of host and reddening still results in higher
values of $R_{uv}$. The host contribution and reddening will effect
the criteria defined with an optical luminosity the most. However,
both the $q_{24}$ and $P_{5GHz}$ criteria are insensitive to
reddening and host contribution, and we find both $q_{24}$ and
$P_{5GHz}$ to agree with $R_L$ for this sample
(Figure~\ref{rlfig1}). This suggests that neither reddening nor
host-contamination are the main reasons for the low RL fraction in
this sample.

\begin{figure}
\epsfig{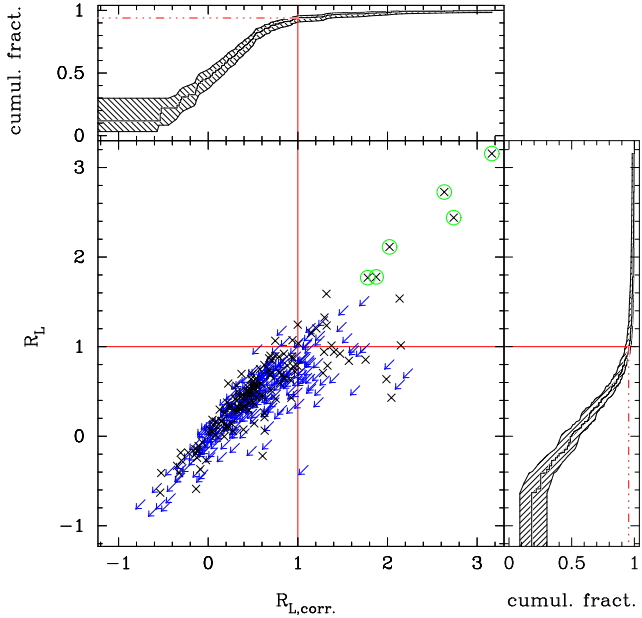}
\caption{The radio-loudness after correction for the host
contribution and reddening ($R_{L,corr}$) compared to the observed
radio-loudness ($R_L$). The plot is color-coded as in
Figure~\ref{rlfig1}\&\ref{rlfig2}. \label{rlcomp}}
\end{figure}

To ascertain in more detail that reddening and host-contamination do
not play a major part in producing the low radio-loud fraction in
the XMM-COSMOS sample, we use the mixing diagram (Hao et al. 2013)
to estimate the host galaxy fraction at $1~\mu m$ ($f_g$) for each
quasar. Here we choose the mixing curve connecting Ell5 (SWIRE
galaxy template of elliptical galaxy with age of 5 Gyr, Polletta et
al. 2007) and E94 (Elvis et al. 1994 mean quasar SED). As shown in
Hao et al. (2013), the difference in galaxy fraction is negligible
regardless of the mixing curves chosen. The AGN contribution of the
luminosity at $1 \mu m$ is thus $L_{AGN, 1\mu m}=L_{1\mu m}(1-f_g)$.
We assume that the un-reddened intrinsic quasar SED has the shape of
the E94 SED, which is a reasonable assumption (Hao et al. 2014). We
normalize the E94 SED to the pure AGN luminosity at $1 \mu m$, and
recover the AGN rest-frame B band luminosity from
$L_{B,corr}=L_{1\mu m}(1-f_g)-L_{E94, 1\mu m}+L_{E94, B}$. As the
radio emission mainly comes from the AGN, the new
$R_{L,corr}=\log(L_{5GHz}/L_{B,corr})$ is the radio-loudness
corrected for the host galaxy contribution and reddening.
Figure~\ref{rlcomp} shows the comparison between the radio-loudness
before and after the correction. The radio-loud fraction for this
corrected radio-loudness is $6.08\%^{+3.18\%}_{-1.75\%}$. We can see
that it is slightly higher than the radio-loud fraction before the
correction, which means the host contribution and reddening in
combination decrease the radio-loud fraction. This is consistent
with the case of nearby Seyferts (Ho \& Peng 2001). However, they
are not the most significant factor leading to the low radio-loud
fraction found in the XMM-COSMOS sample.

RL quasars usually reside in very massive galaxies and typically
have a lower optical or X-ray output at given stellar mass (i.e.
lower $L/L_{Edd}$ at given $L$, Sikora et al. 2007) compared to RQ
quasars. This means that a $L_X$-limited sample will have a lower RL
quasar fraction, compared to a mass-limited sample. COSMOS is a deep
survey in a limited area, thus it will sample a few high mass
($L>5L_*$) galaxies at any redshift, compared to wide area shallow
surveys (e.g. SDSS), similar to the CDFN sample (Donley et al. 2007)
which also has a low radio-loud fraction.

We also note that XC407 is not a complete X-ray selected sample, but
a sub-sample with spectroscopic coverage. The optical-near infrared
magnitudes of the spectroscopic sub-sample thus tend to be brighter
than the complete sample, which in principle produces a bias toward
low radio-loudness in the radio-loudness distribution of the
complete sample.

\subsection{Correlation of Radio-loudness with Other Properties}
\begin{figure*}
\includegraphics[angle=0,width=0.32\textwidth]{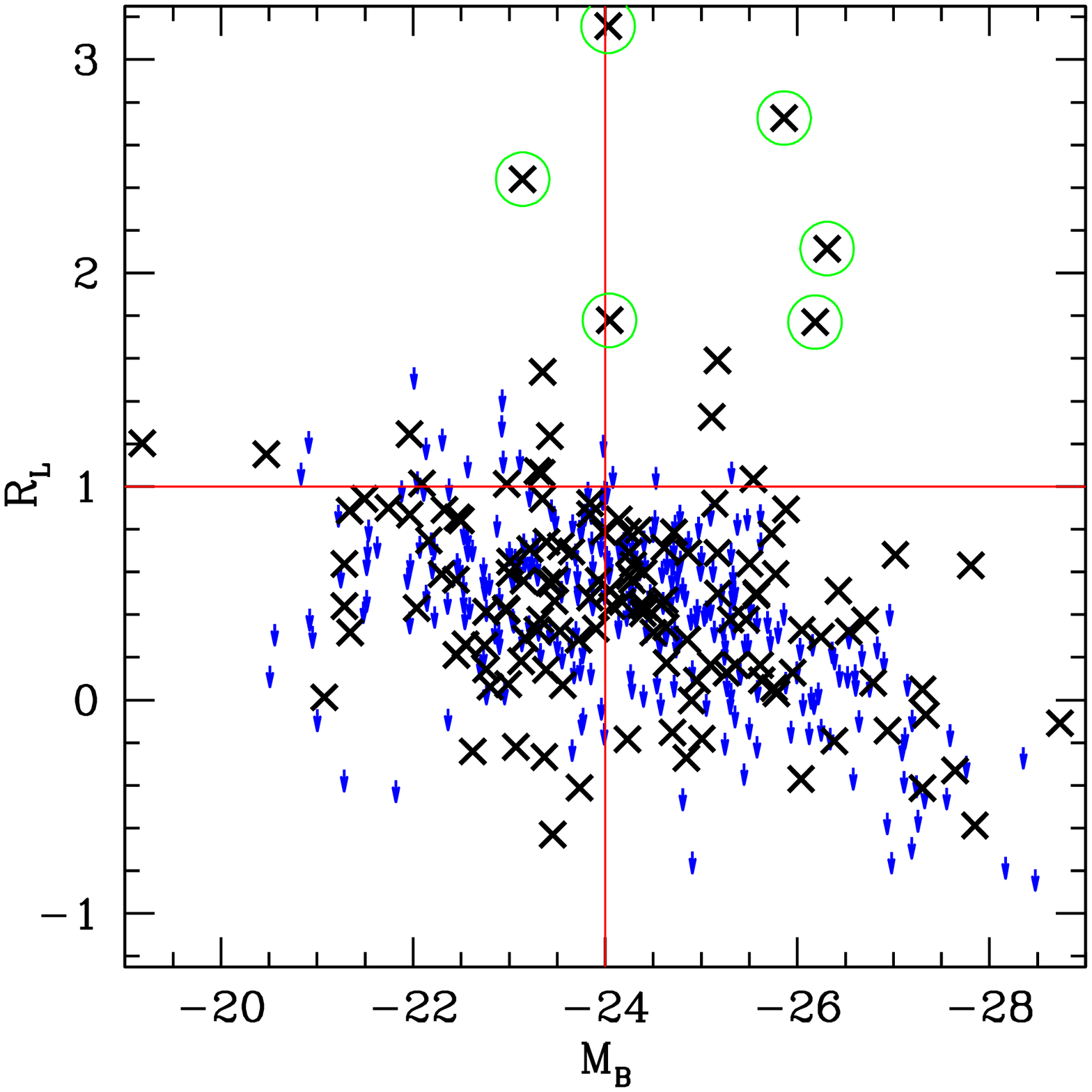}
\includegraphics[angle=0,width=0.32\textwidth]{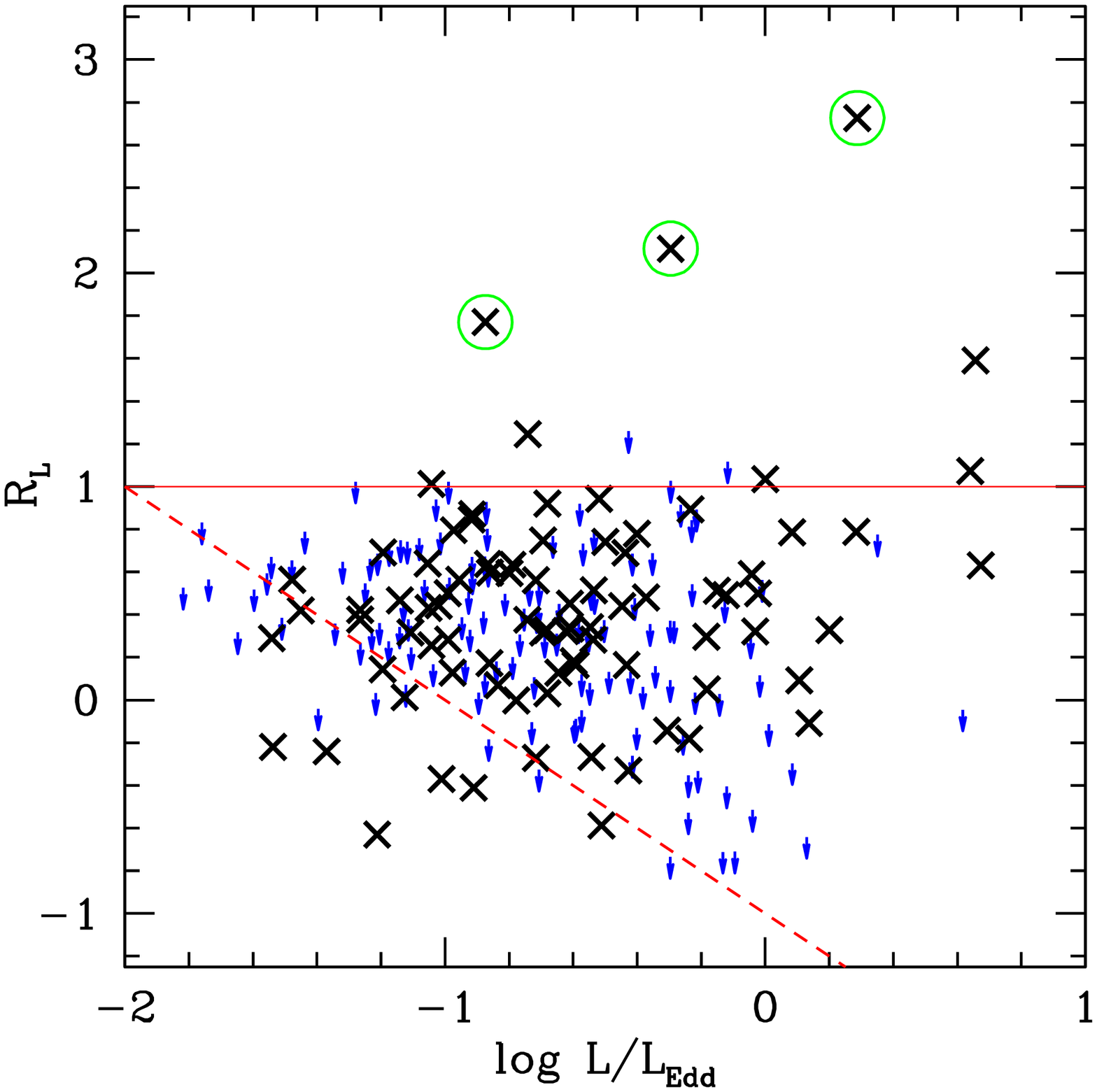}
\includegraphics[angle=0,width=0.32\textwidth]{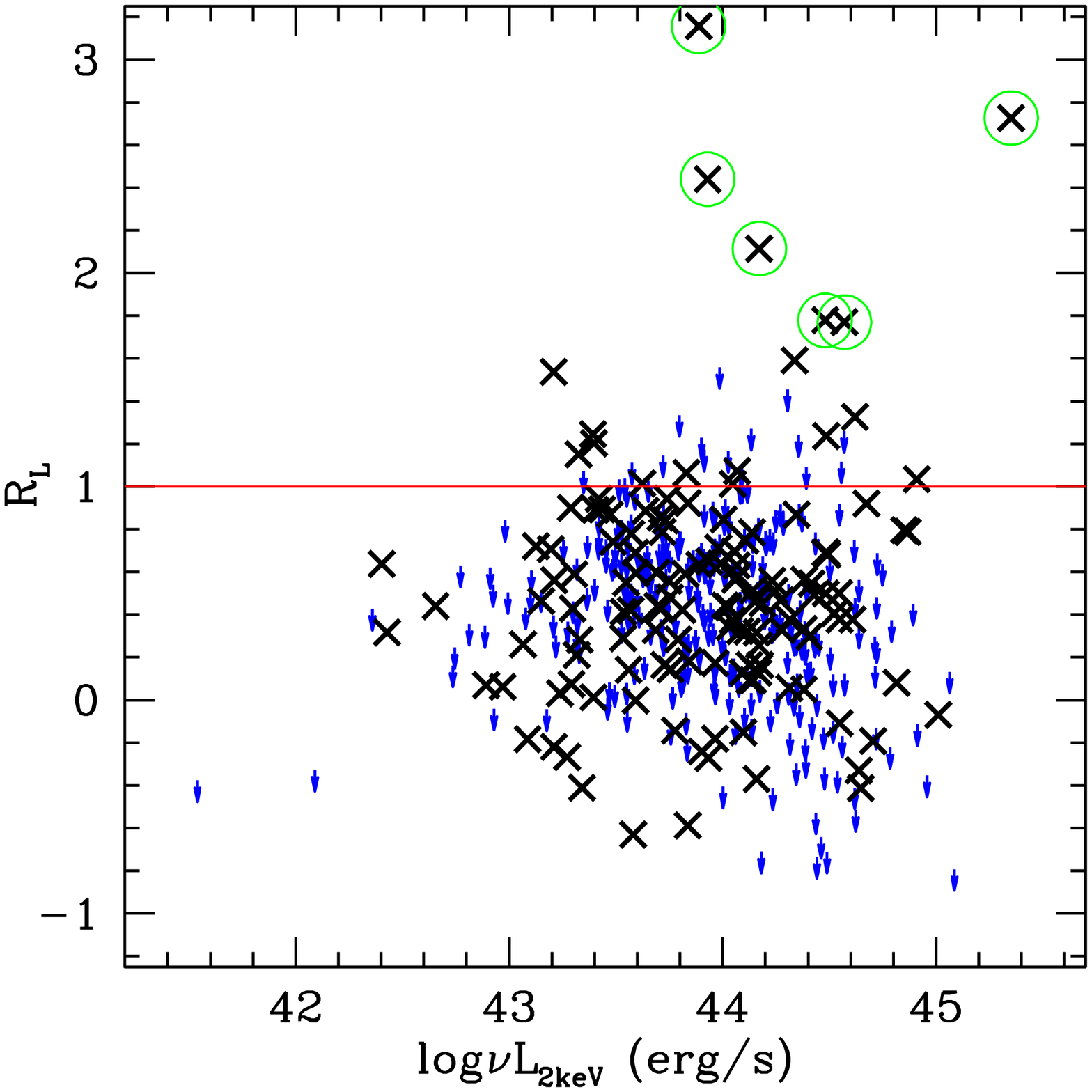}
\caption{$R_L$ versus $M_B$ (left), $\log(L/L_{Edd})$ (middle) and
$\nu L_{2keV}$ (right)  for the XC407 sample, respectively. Here
$M_B$ is calculated using H$_0$ =50~km~s$^{-1}$~Mpc$^{-1}$. The
Eddington ratio is $\lambda_{Edd}=L_{bol}/L_{Edd}$. The $\nu
L_{2keV}$ is the 2keV X-ray luminosity in units of erg/s. Note that
in the middle plot, we only show 204 among the 407 quasars in the
XC407 that have black hole mass estimates (Trump et al. 2009b;
Merloni et al. 2010). For the rest of the sample, the broad emission
lines are locate at the edge of the spectrum, such that reliable
estimates of the black hole mass are not possible (Elvis et al.
2012). The black crosses show the quasars with $>3\sigma$ detections
in the radio and the blue arrows show the quasars with upper limits
in the radio. The green circles indicate the 6 sources classified as
RL by all criteria. The red dashed line in the center panel shows
$R_L=-\log(L/L_{Edd})-1$ which represents the general trend found in
Sikora et al. (2007). \label{rlcomp}}
\end{figure*}

In the EMSS, Della Ceca et al. (1994) found a 10.6\% radio-loud
fraction and a trend of lower RL fractions for absolute magnitudes
fainter than M$_B$ = - 24. Note that this sample is also X-ray
selected but is not restricted to type 1 AGNs. The XMM-COSMOS B band
absolute magnitude distribution is similar to that of the EMSS
sample (Elvis et al. 2012), with 191 (46\% of the whole sample)
having M$_B>$ - 24. We plot $R_L$ versus $M_B$ for XC407 in the left
panel of Figure~\ref{rlcomp}. Note that we have changed to the same
cosmology (H$_0$ =50~km~s$^{-1}$~Mpc$^{-1}$) as Della Ceca et al.
(1994) for $M_B$. No trend of reduced RL fraction at $M_B<-24$ is
seen in our sample.

Radio-loudness is seen to increase with decreasing Eddington ratio
($L/L_{Edd}$), in particular $R_L\propto(L/L_{Edd})^{-1}$ at
L/L$_{Edd}>0.001$ (Sikora et al. 2007, Ho 2002). However, the
XMM-COSMOS sources have relatively high Eddington ratios ($>0.01$,
with a median of 0.2, Hao et al. 2014). This is similar to the
values for PG quasars (Kellerman et al., 1987, Sikora et al. 2007).
We plot $R_L$ versus $\log(L/L_{Edd})$ in the middle panel of
Figure~\ref{rlcomp}. No trend of reduced RL fraction for large
Eddington ratio is visible in our sample. In fact, all the $R_L>1$
AGNs have $\log(L/L_{Edd})>-1$.

In E94, the mean SED of the RL and RQ sample show differences in the
X-rays (see Figure~\ref{rlsed} red and green curve), with RL quasars
tending to have relatively brighter X-ray luminosity (about 0.5 dex
higher). We checked the $R_L$ versus the X-ray luminosity at 2keV
($\nu L_{2keV}$) in the right panel of Figure~\ref{rlcomp}. No
obvious trend toward reduced RL fraction at low X-ray luminosity is
seen. If we divided the sample at the median $\nu L_{2keV}$ (43.96
erg/s), slightly more radio loud quasars are above the median, which
is 10 radio-loud quasars with radio detections has X-ray luminosity
above median compared to 8 radio-loud quasars with radio detections
has X-ray luminosity below the median. This is consistent with the
mean SED get in E94.

\section{Summary}

We have used nine different radio-loudness criteria to study the
radio loud fraction of the XMM-COSMOS type 1 AGN sample, in which
six were defined in the rest-frame (radio loud fraction ranging from
1.5\%--4.5\%) and three were defined in the observed frame without
k-correction. The poor statistics on the RL sources does not allow
to infer statistically significant results on a dichotomy between RQ
and RL AGN, of which we do not see any sign.

The criteria defined in the rest-frame generally agree with each
other and gives similar radio-loud fractions. The criterion $R_X$
and $P_{5GHz}$ agree the best with the smallest number of ambiguous
sources. Radio-loudness defined via a radio-to-optical luminosity
ratio ($R_L$, $R_{uv}$ and $R_i$) display the strongest correlation.
The radio power ($P_{5GHz}$) gives the strongest restriction, that
yields the lowest radio-loud fraction for any single criteria. If we
require all the criteria to be satisfied at the same time, the
radio-loud fraction is marginally smaller than the radio-loud
fraction we get from the $P_{5GHz}$ criterion only.

Two of the criteria defined in the observed frame without
k-correction give a much higher radio-loud fraction, but if we take
the redshift distribution of the sample into consideration, the
radio-loud fraction is greatly reduced and becomes consistent with
the results from the criteria defined in the rest-frame. Thus, we
need to be careful when citing the radio-loud fraction using the
criteria defined in the observed frame without k-correction.

If we corrected the host galaxy contribution and reddening, the
radio-loud fraction in the XC407 sample will rise to
$6.08\%^{+3.18\%}_{-1.75\%}$, still a bit smaller than the typical
value of 10\% in optical selected samples, which might be caused by
the selection of the sample being $L_X-$limited and biasing towards
bright optical quasars. No correlation of the radio-loudness with
$M_B$, $L/L_{Edd}$ or $L_X$ is seen in the XC407 sample.

The combination of newly approved deep 3GHz EVLA observations (P.I.:
Smolcic) with the completion of Chandra coverage of the whole COSMOS
field in the COSMOS Legacy survey (P.I.: Civano) will help us
understand the origin of the low RL fraction of type 1 AGNs.

\section*{Acknowledgments}

HH thank Belinda Wilkes for useful discussion. This work was
supported in part by NASA {\em Chandra} grant number G07-8136A (HH,
ME, CV). LCH acknowledges support from the Kavli Foundation, Peking
University, and the Chinese Academy of Science through grant No.
XDB09030100 (Emergence of Cosmological Structures) from the
Strategic Priority Research Program. Support from the Italian Space
Agency (ASI) under the contracts ASI-INAF I/088/06/0 and I/009/10/0
is acknowledged (AC and CV). MS acknowledges support by the German
Deutsche Forschungsgemeinschaft, DFG Leibniz Prize (FKZ HA
1850/28-1). KS gratefully acknowledges support from Swiss National
Science Foundation Grant PP00P2\_138979/1.


\label{lastpage}

\end{document}